\documentclass[smallextended]{svjour3}

\smartqed

\usepackage{natbib}

\usepackage{algorithm}
\usepackage{algpseudocode}
\usepackage{eurosym}
\usepackage{float}
\usepackage{amssymb}
\usepackage{amsmath}
\usepackage{graphicx}
\usepackage[T1]{fontenc}
\usepackage[utf8]{inputenc}
\usepackage{flushend}
\usepackage{chngcntr}
\usepackage{url}
\usepackage{xspace}
\usepackage{listings}
\usepackage{fancyhdr}
\usepackage[normalem]{ulem}
\usepackage{subfig}
\usepackage{soul}
\usepackage{enumerate}
\usepackage[shortlabels]{enumitem}
\usepackage{soulutf8}

\sethlcolor{lightgray}

\usepackage{hyperref}
\hypersetup{hidelinks}

\usepackage{multirow}
\usepackage[table,xcdraw]{xcolor}

\newcommand{\orcid}[1]{\href{https://orcid.org/#1}{\includegraphics[width=10pt]{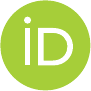}}}

\usepackage{geometry}

\usepackage[misc,geometry]{ifsym}
\newcommand*{\affmark}[1][*]{\textsuperscript{#1}}

\renewcommand{\cite}[1]{\citep{#1}}

\journalname{Empirical Software Engineering}

\begin{document}

\title{OpenSCV: An Open Hierarchical Taxonomy for\\Smart Contract Vulnerabilities}


\author{
Fernando Richter Vidal\affmark[1]\orcid{0000-0003-4869-2336}
\and
Naghmeh Ivaki\affmark[1]\orcid{0000-0001-8376-6711}
\and
Nuno Laranjeiro\affmark[1]\orcid{0000-0003-0011-9901}
}

\authorrunning{F. R. Vidal \and N. Ivaki \and N. Laranjeiro}

\institute{Fernando Richter Vidal\\
fernandovidal@dei.uc.pt\\
\\
            Naghmeh Ivaki\\
              naghmeh@dei.uc.pt\\
              \\
           Nuno Laranjeiro\\
              cnl@dei.uc.pt\\
              \\
\affmark[1] University of Coimbra, Centre for Informatics and Systems of the University of Coimbra, Department of Informatics Engineering, Portugal
}


\date{Received: date / Accepted: date}

\maketitle

\begin{abstract}

Smart contracts are nowadays at the core of most blockchain systems, as they specify and allow an agreement between entities that wish to perform a transaction. As any computer program, smart contracts are subject to the presence of residual faults, including severe security vulnerabilities, which require that the vulnerable contract is terminated in the blockchain. In this context, research began to be developed to prevent the deployment of smart contract holding vulnerabilities, mostly in the form of vulnerability detection tools. Along with these efforts, several and heterogeneous vulnerability classification schemes arised (e.g., most notably DASP and SWC). At the time of writing, these are mostly outdated initiatives, despite the fact that smart contract vulnerabilities are continuously being discovered and the associated rich information being mostly disregarded. In this paper, we propose OpenSCV, a new and Open hierarchical taxonomy for Smart Contract Vulnerabilities, which is open to community contributions and matches the current state of the practice, while being prepared to handle future modifications and evolution. The taxonomy was built based on the analysis of research on vulnerability classification, community-maintained classification schemes, and research on smart contract vulnerability detection. We show how OpenSCV covers the announced detection ability of current vulnerability detection tools, and highlight its usefulness as a resource in smart contract vulnerability research.

\end{abstract}

\keywords{
Blockchain \and Smart contract \and Taxonomy \and Vulnerability \and Classification \and Software Fault; Software Defect}


\section{Introduction}
\label{sec:intro}

Smart contracts play an important role in advancing blockchain because they expand the application of the technology to various domains (e.g., education \cite{Grech2017a}, healthcare \cite{Zhang2022}, government \cite{Geneiatakis2020}). While they are essential for the consolidation and expansion of the technology, they also bring serious risks, namely those associated with the potential presence of vulnerabilities that can affect the security of the blockchain system \cite{Atzei2017}.

Just as conventional programs, smart contracts are being deployed with residual software faults (i.e., bugs or defects), including security vulnerabilities (i.e., internal faults that enable external events to harm the system) \cite{Qian2022, BasicConcepts2004}. However, the consequences of deploying a faulty contract have particular characteristics in the context of blockchain systems, such as: i) if faulty code is identified, the respective contract cannot be patched, it must be terminated, and a new one should be created \cite{8847638}; ii) once the potentially erroneous data (generated/updated by faulty contracts) has been stored in the blockchain, there is no way to change it, i.e., to undo the respective transactions (and subsequent transactions that rely on this data) \cite{Yaga2018}; and iii) if the faulty contract has been executed, the associated impact may be irreparable (e.g., reputation costs) \cite{antonopoulos_mastering_2018}.

Several initiatives have been created that ultimately aim at contributing to the development of more secure smart contracts.  Among these initiatives, we find three main types: i) New smart contract programming languages (e.g.,  Clarify  \cite{Blockstack2021}, Vyper \cite{Vyper.vs.solidity.2020}, Obsidian \cite{Coblenz2019}), which aim at increasing protection against faults; ii) New vulnerability detection tools (e.g., Mythril \cite{Mythril}, Neucheck \cite{lu_neucheck_2019}, \cite{SAILFISH_2022}, SoliDetector \cite{Hu2023}), which have the main goal of detecting vulnerabilities in smart contracts so that vulnerable contracts do not reach the deployment phase; and also, iii) and also vulnerability classifications that mostly allow knowledge regarding vulnerabilities to be identified in a standard manner and systematized.

The existence of vulnerability (or software defects, in general) classifications is quite important, as we can observe by the research and industry effort associated with well-known cases like OWASP \cite{OWASP2001}, NVD \cite{U.S.government1999}, CVE \cite{CVE1999}, CWE \cite{CWECommunity2009}, or, in the case of smart contracts most notably by SWC \cite{swc}, and DASP \cite{nccgroup_decentralized_2021}. Generally, they raise the level of awareness among developers and may allow, in a uniform manner, for development tools to assist developers regarding mistakes being placed in the code. They may also help in the design and development of vulnerability detection tools and in the assessment of their detection capabilities \cite{durieux_empirical_2020, Hu2021, di_angelo_survey_2019}. This case also holds for programming languages. It is known that languages, such as Obsidian, have benefited from the systematized knowledge of vulnerabilities. There are even studies that use taxonomies as a basis for comparing different programming languages with respect to the protection offered against certain types of vulnerabilities, e.g., \cite{Vyper.vs.solidity.2020}.

At the time of writing, vulnerability classifications for smart contracts have significant limitations. They are generally \textbf{outdated} with, at the time of writing, popular schemes like DASP or SWC not being updated since 2018. This largely differs from the state of the practice, in which we find cases of tools like Securify2 \cite{Tsankov2018} already detecting several vulnerabilities for which there is no accurate description. As in other software areas, with new vulnerabilities being continuously discovered, having a flexible way of integrating (and possibly restructuring the classification) new defects is crucial.

Vulnerability naming and classification schemes are being defined using \textbf{arbitrary nomenclatures}. This is easily visible just by analyzing a few of the most cited papers in vulnerability detection, e.g., \cite{luu_making_2016,tsankov_securify_2018,kalra_zeus_2018}. The lack of a standard nomenclature leads to verification tools mostly using arbitrary names to present their result, e.g., SmartCheck \cite{tikhomirov_smartcheck_2018} and Slither \cite{feist_slither_2019} respectively use \textit{balance equality} and \textit{incorrect-equality/locked-ether} to refer to the same vulnerability. As a result, it is very difficult to compare the effectiveness of different tools. As classifications are many times built based on multiple sources, such as different industry tools and several research papers \cite{Rameder2022}, terms easily end up being inconsistent. This is aggravated when there is no active maintenance even for known issues. Indeed, \textbf{Reduced community contribution} is known to be a problem, with the main classifications that are community-oriented (i.e., DASP, SWC) showing residual community activity, many times related to minor issues (e.g., broken links) \cite{DASP,swc}.

Many times, vulnerability classification schemes mix the characteristics of a certain vulnerability with the effect of exploiting it, how it is exploited, or its impact. This \textbf{concept inconsistency} is quite visible in current taxonomies. As an example, in \cite{Vyper.vs.solidity.2020} presents \textit{DoS with unbounded operation} as a vulnerability, but it is not possible to understand what the vulnerability is with this name (e.g., it can be a problem in a loop, it can be a malicious call that is externally triggered several times). Instead, the given name refers to the possible impact of exploiting a vulnerability, which should be a separate dimension for characterizing the defect. Similarly, this occurs in DASP \cite{DASP}, in which one of the categories is precisely \textit{Denial of Service}. Another aspect this latter example shows is that taxonomies are being built with \textbf{inadequate granularity}, often too coarse to be really helpful. For instance, the \textit{Denial of Service} category in DASP may refer to \textit{gas limit reached}, \textit{unexpected throw}, \textit{unexpected kill}, or \textit{access control breached}. Moreover, the description is sometimes so short that may become ambiguous (e.g., \textit{access control breached} may refer to a vulnerability that would simply fit in \textit{access control}, which is another DASP category).

\textbf{In this paper, we propose OpenSCV, a new hierarchical and Open taxonomy for Smart Contract Vulnerabilities} (available at \url{http://openscv.dei.uc.pt}), which is open to community contributions \cite{openscvGithub}, aims at matching the current state of the practice and is prepared to handle future modifications and evolution. To build the taxonomy, we analyzed current smart contract vulnerability classifications and discussed their gaps and limitations. We then analyzed the announced detection capabilities of 49 research works on smart contract vulnerability detection with the goal of collecting an heterogeneous set of 357 vulnerability definitions. We then mapped the vulnerabilities in existing classifications, namely DASP \cite{DASP}, SWC \cite{swc}, \cite{Rameder2022}, and CWE \cite{CWECommunity2009} and further characterized them using the Orthogonal Defect Classification (ODC) \cite{ODC,odcExtension} and with a code excerpt. Names were then consolidated and grouped in a structure that was built bottom-up. This process involved 2 Experienced Researchers and 1 Early Stage Researcher, which revised the proposed taxonomy iteratively in terms of structure, correctness, and uniformity.

We structured OpenSCV to allow it to be flexible to changes and evolution by preparing a supporting infrastructure at github. We are able to receive change requests easily and integration information from new research on vulnerability detection into the taxonomy. All OpenSCV entries are supported by a code example, with the goal of mitigating possible ambiguities in the description of each vulnerability and we also prepared an initial dataset holding vulnerable contracts (one per each of the vulnerabilities present in OpenSCV) and their respective correction. OpenSCV is live and available at \url{http://openscv.dei.uc.pt} \cite{openscvSite}, the github repository is available at \cite{openscvGithub} and linked to Zenodo which permanently hosts the dataset \cite{openscvZenodo}. It is worthwhile mentioning that the taxonomy considers mostly software vulnerabilities and a few software defects considered in the literature to be associated with high-security risks. For simplicity, \textit{we use the term vulnerability throughout the paper} to refer to both cases.

The rest of this paper is organized as follows. section 2 discusses the related work and limitations of current vulnerability classification schemes. Section 3 presents the process followed to build the taxonomy and overviews the final outcome. Section 4 presents the taxonomy structure and provides a brief description of all vulnerabilities included in the taxonomy. Section 5 characterizes and discusses the coverage of the taxonomy in perspective with the state-of-the-art. Section 6 presents the threats to the validity of this work, and finally, Section 7 concludes this paper.
\section{State of the Art}
\label{sec:SectionII}

This section presents the quality properties of taxonomies and then discusses the existing classification schemes for smart contract vulnerabilities. The classification schemes presented are have their origins in: a) existing research on smart contract vulnerability classification; b) community-oriented initiatives; and also c) vulnerability detection research. The section closes with a discussion of the gaps and limitations of current classification.

\subsection{Taxonomy Quality Properties}

We analyzed a set of reference works centered around the definition of taxonomies as well as critical analyses of vulnerability taxonomies \cite{Bishop1996, lindqvist_how_1997, Mann1999, Rameder2022, lough_taxonomy_2001, hansman_taxonomy_2005} to identify a set of quality properties criteria, which should be followed when designing a taxonomy that is expected to be long-lived. The following paragraphs discuss the identified properties.

A classification system may benefit from a \textbf{hierarchical organization} as it allows to show similar characteristics of related vulnerabilities, which may be helpful for vulnerability prevention \cite{Bishop1996}. A hierarchical structure may be a tree in which each node refers to a category of vulnerabilities, and each leaf corresponds to individual vulnerabilities. Thus, the granularity of the categories should generally vary from large to fine as we traverse the tree from the root to the leaves.

Nodes at a certain tree level must be as uniform as possible, i.e., ideally representing the same \textbf{level of abstraction} or a group of vulnerabilities viewed from the same perspective. Obviously, this is quite difficult to achieve because, many times, this has to be balanced with the creation of taxonomy trees that become too complex, which in the end, may make it less comprehensible or less helpful. Also, sometime the nature of the problem is simply an unbalanced or heterogeneous (in structure) one, which basically disallows this criteria. Anyway, a uniform taxonomy may contribute to fewer errors (in its use) and, as such, a higher probability of adoption by practitioners. In practice, it may contribute to a taxonomy that is \textbf{useful} and \textbf{comprehensible} \cite{lindqvist_how_1997} (i.e., understandable by security experts but also by less specialized people). 

The selection of names to be used in a classification scheme is particularly important. The name that describes a certain vulnerability must be a \textbf{unique identifier}) and \textbf{non-ambiguous} \cite{howard_analysis_1997, lindqvist_how_1997}, meaning that the name and also the associated description must allow not only for easy identification but should include enough information to distinguish it from other vulnerabilities \cite{Mann1999, Bishop1996}. Whenever possible, existing \textbf{terminology} should be used \cite{lindqvist_how_1997}. The name and characteristics of a certain defect should characterize what the issue is and not additional dimensions, such as the effect of exploiting it. While it is acceptable to understand the effect of exploiting a certain vulnerability starting from its description, the characteristics of the problem itself cannot be omitted and should be clearly identified \cite{Mann1999, Bishop1996}.

Regardless of the perspective of the individual using the taxonomy, a certain defect should be classified in the same manner by different individuals (e.g., developers, users, testers). This means that not only the names and structure should be as clear as possible, but also that the process of classifying a certain defect must be made clear (whenever the structure and nomenclature are not sufficient), i.e., there must be a \textbf{deterministic} \cite{krsul_software_1998} way of classifying a certain defect, which fosters \textbf{repeatability} \cite{howard_analysis_1997, krsul_software_1998} of using the classification.

Finally, a taxonomy should also allow for \textbf{completeness} \cite{amoroso_fundamentals_1994}, i.e., the taxonomy should provide a \textbf{good coverage} \cite{Rameder2022} of the vulnerabilities identified in state of the art or reported by vulnerability detection tools. Also, it should be \textbf{open to the community} (i.e., accept new entries from the community) and shareable (i.e., no distribution restrictions) \cite{Mann1999}. The fact that it is open is also a factor that can contribute to it being \textbf{accepted} \cite{amoroso_fundamentals_1994, howard_analysis_1997}.


\subsection{Smart Contract Vulnerability Classification Schemes}

To the best of our knowledge, the first initiative to classify smart contract vulnerabilities (for Ethereum systems) is proposed in \cite{Atzei2017}. The authors listed 12 vulnerabilities, which we overview in Table \ref{tab:atzei}, and implemented nine of the corresponding attacks.

\begin{table}[ht]
\scriptsize
\centering
\caption{Classification proposed in \cite{Atzei2017}.}
\label{tab:atzei}
\begin{tabular}{|l|l|}
\hline
\textbf{Level}                                    & \textbf{Vulnerability} \\ \hline
{Solidity}                         & Call to the unknown     \\ \cline{2-2} 
                                                  & Gasless send           \\ \cline{2-2} 
                                                  & Exception disorders    \\ \cline{2-2} 
                                                  & Type casts             \\ \cline{2-2} 
                                                  & Reentrancy             \\ \cline{2-2} 
                                                  & Keeping secrets        \\ \hline
{EVM}                              & Immutable bugs         \\ \cline{2-2} 
                                                  & Ether lost in transfer \\ \cline{2-2} 
                                                  & Stack size limit       \\ \hline
\multicolumn{1}{|c|}{{Blockchain}} & Unpredicable state     \\ \cline{2-2} 
\multicolumn{1}{|c|}{}                            & Generating randomness  \\ \cline{2-2} 
\multicolumn{1}{|c|}{}                            & Time constraints       \\ \hline
\end{tabular}
\end{table}

This initial effort is quite relevant but holds some limitations. Some of the selected names do not really specify the nature of the vulnerabilities or are not clear about the problem being characterized (e.g., "call to the unknown"). This limitation was mitigated in \cite{Zhou2022} and \cite{arganaraz_detection_2020}, where the authors tried to make the names used more specific. In \cite{Atzei2017} three categories of issues are proposed: i) Solidity Issues (i.e., language weaknesses), ii) EVM Issues (i.e., residuals faults in byte code), and iii) Blockchain Issues (i.e., vulnerabilities from blockchain technology). Despite allowing an initial separation of the issues (which may help developers in dealing with the faults), this scheme does not benefit from the presence of a more complex hierarchy, which is a better fit for cases where we find several interrelated families of vulnerabilities. We have also identified that these three categories may generate some ambiguity as some cases could potentially fit into multiple categories. For example, \textit{Immutable Bugs} could be classified into EVM or Blockchain. Despite this, the separation between the cases referring to the programs (i.e., solidity source code or EVM binary code) and the blockchain platform is helpful. This classification is not available in a public repository and, due to its age, its coverage is relatively low, accounting for 12 evulnerabilities.

Table  \ref{tab:Vyper} overviews the vulnerability classification presented in \cite{Vyper.vs.solidity.2020},  which has the goal of allowing comparison between the security of the Solidity and Vype languages. The work presents 18 vulnerabilities, along with a detailed explanation for each one, and individual code examples for each vulnerability. Being mostly a list of vulnerabilities, there are no benefits associated with hierarchical structures. There is no open public repository associated with the proposal, and the 18 vulnerabilities are nowadays a small amount, e.g., the work in \cite{ Rameder2022} identifies a total of 54 defects.

 \begin{table}[ht]
\scriptsize
\centering
\caption{Classification proposal in \cite{Vyper.vs.solidity.2020}.}
\label{tab:Vyper}
\begin{tabular}{|l|}
\hline
\textbf{Vulnerabilities}             \\ \hline
Integer   overflow and underflow     \\ \hline
DoS with   unbounded operation       \\ \hline
Unchecked call   return value        \\ \hline
Reentrancy                           \\ \hline
Delegate call   injection            \\ \hline
Forced Ether to contract             \\ \hline
DoS with unexpected revert           \\ \hline
Erroneous visitility                 \\ \hline
Uninitialized storage pointer        \\ \hline
Upgradeable contract                 \\ \hline
Type casts                           \\ \hline
Insufficient signature   information \\ \hline
Frozen Ether                         \\ \hline
Authentication through tx.   Origin  \\ \hline
Unprotected suicide                  \\ \hline
Leaking Ether to arbitrary   address \\ \hline
Secrecy failure                      \\ \hline
Outdated compiler version            \\ \hline
\end{tabular}
\end{table}

A vulnerability classification is presented in \cite{arganaraz_detection_2020} with the goal of exposing threats and, ultimately, minimizing the presence of software bugs in smart contracts. Table \ref{tab:arganaraz} presents the proposed classification in which we find faults separated into two levels: i) security (i.e., faults that may be exploited by attacks; and ii) functional (i.e., faults that violate the program's functionality). Each fault is also associated with a criticality level, which may be useful for getting developers' attention while coding. We found that certain cases, such as \textit{Non-verified maths} and \textit{Malicious libraries}, may indicate the same vulnerability, indicating a potential need for further clarification and refinement of the classification process to address any ambiguity. Similar to the previously presented works, there is no hierarchical structure besides the two groups of vulnerabilities. The names used in the classification are quite specific (i.e.,  \textit{Use of tx.origin}), which makes it difficult to understand the problem in a more abstract manner. Only 13 faults are considered, with no possibility of expansion. Still, the idea of classifying the faults into two broad concepts of security and functionality is a vision that may be interesting for newer classifications (e.g., targeting different types of systems).

\begin{table}[ht]
\scriptsize
\centering
\caption{Classification proposal in\cite{arganaraz_detection_2020}.}
\label{tab:arganaraz}
\begin{tabular}{|l|l|l|}
\hline
\textbf{Level}              & \textbf{Vulnerability}                              & \textbf{Impact}  \\ \hline
{Security}   & Equality on the balance                             & Average \\ \cline{2-3} 
                            & Non-verified external call                          & High    \\ \cline{2-3} 
                            & Use of send instead of transfer                     & Average \\ \cline{2-3} 
                            & Denial of a service because of an external contract & High    \\ \cline{2-3} 
                            & Re-entrancy                                         & High    \\ \cline{2-3} 
                            & Malicious libraries                                 & Low     \\ \cline{2-3} 
                            & Use of tx.origin                                    & Average \\ \cline{2-3} 
                            & Transfer of all the gas                             & High    \\ \hline
{Functional} & Integer division                                    & Low     \\ \cline{2-3} 
                            & Blocked money                                       & Average \\ \cline{2-3} 
                            & Non-verified maths                                  & Low     \\ \cline{2-3} 
                            & Dependence on the timestamp                         & Average \\ \cline{2-3} 
                            & Unsecure inference                                  & Average \\ \hline
\end{tabular}
\end{table}

A smart contract vulnerability classification is presented in \cite{Zhou2022}, based on the previous work in \cite{Atzei2017}. In this classification, summarized in Table \ref{tab:Zhou}, the groups were maintained (i.e., Solidity, EVM, Blockchain), but the vulnerability entries were modified (i.e., some names were removed, like \textit{stack size limit} and \textit{gasless send} and others names were included, like \textit{tx. origin} and \textit{default visibility}). The authors linked the proposed names to an external taxonomy, namely CWE \cite{CWECommunity2009}, which is helpful for understanding each vulnerability, verifying the correctness of the proposed classification, and also for standardization purposes. The proposed classification defines a basic separation of vulnerabilities, mostly distinguishing cases related to the programs from cases related to the platform. Again the number of vulnerabilities listed is quite small (i.e.,13 vulnerabilities), and the work could benefit from a repository open to community contributions.

\begin{table}[ht]
\scriptsize
\centering
\caption{Vulnerability classification in \cite{Zhou2022}.}
\label{tab:Zhou}
\begin{tabular}{|c|l|l|l|}
\hline
\multicolumn{1}{|l|}{\textbf{Level}} & \textbf{Vulnerability}                                                        & \textbf{CWE} & \textbf{Real-Word Attack}                                                           \\ \hline
{Solidity}            & Re-entrancy                                                                   & CWE-841      & \begin{tabular}[c]{@{}l@{}}The   DAO\\      Attack\end{tabular}                     \\ \cline{2-4} 
                                     & Arithmetic issues                                                             & CWE-682      & \begin{tabular}[c]{@{}l@{}}PoWHcoin\\      attack\end{tabular}                      \\ \cline{2-4} 
                                     & \begin{tabular}[c]{@{}l@{}}Delegatecall to \\ insecure contracts\end{tabular} & CWE-829      & \begin{tabular}[c]{@{}l@{}}Parity\\      Wallet\\ (Second\\      Hack)\end{tabular} \\ \cline{2-4} 
                                     & Selfdestruct                                                                  & CWE-284      & \begin{tabular}[c]{@{}l@{}}Parity   Library\\      bug\end{tabular}                 \\ \cline{2-4} 
                                     & Tx.origin                                                                     & CWE-477      & -                                                                                   \\ \cline{2-4} 
                                     & Mishandled exception                                                          & CWE-252      & \begin{tabular}[c]{@{}l@{}}King of The \\ Ether attack\end{tabular}                 \\ \cline{2-4} 
                                     & Default visibility                                                            & CWE-710      & \begin{tabular}[c]{@{}l@{}}Parity\\      Wallet\\ (First\\      Hack)\end{tabular}  \\ \cline{2-4} 
                                     & External contract referencing                                                 & CWE-829      & Honey   Pot                                                                         \\ \hline
{EVM}                 & \begin{tabular}[c]{@{}l@{}}Short   address/parameter\\  issues\end{tabular}   & CWE-88       & -                                                                                   \\ \cline{2-4} 
                                     & Freezing Ether                                                                & CWE-17       & -                                                                                   \\ \hline
{Blockchain}          & Transaction order   dependence                                                & CWE-362      & \begin{tabular}[c]{@{}l@{}}Attack   on\\      Bancor\end{tabular}                   \\ \cline{2-4} 
                                     & Generating randomness                                                         & CWE-330      & \begin{tabular}[c]{@{}l@{}}PRNG\\      contract\end{tabular}                        \\ \cline{2-4} 
                                     & Timestamp dependence                                                          & CWE-829      & \begin{tabular}[c]{@{}l@{}}GovernMental\\      attack\end{tabular}                  \\ \hline
\end{tabular}
\end{table}

Table \ref{tab:Amiet} presents a vulnerability classification proposed by \cite{Amiet2021}. The classification is based in two categories: i) core blockchain vulnerabilities (i.e., vulnerabilities related to the blockchain platform); ii) smart contracts vulnerabilities (i.e., vulnerabilities related to the programs deployed in the blockchain). At the blockchain level, examples are provided (e.g., attacks on the consensus mechanism), whereas, at the contract level, pseudo-code is presented, which clarifies the security issues identified. These two broad groups are a basis for applying the classification to other types of systems. There are no further hierarchical levels present in this taxonomy, and we found vulnerability names that are unclear, such as \textit{Improper Blockchain Magic Validation}, which does not really characterize the technical details involving the vulnerability. As with previous cases, the 12 vulnerabilities represent a quite small number of currently known vulnerabilities.

\begin{table}[ht]
\scriptsize
\centering
\caption{Classification proposed in \cite{Amiet2021}.}
\label{tab:Amiet}
\begin{tabular}{|l|l|}
\hline
\textbf{Group}                     & \textbf{Vulnerabilities}                \\ \hline
{Core   Blockchain} & Consensus Mechanism   Manipulation      \\ \cline{2-2} 
                                   & Underlying Cryptosystem Vulnerabilities \\ \cline{2-2} 
                                   & Improper Blockchain Magic Validation    \\ \cline{2-2} 
                                   & Improper Transaction Nonce Validation   \\ \cline{2-2} 
                                   & Denial of Service                       \\ \cline{2-2} 
                                   & Public-key and Address Mismatch         \\ \hline
{Smart   Contract}  & Reentrancy                              \\ \cline{2-2} 
                                   & Arithmetic Issues                       \\ \cline{2-2} 
                                   & Unprotected Selfdestruct                \\ \cline{2-2} 
                                   & Visibility Issues                       \\ \cline{2-2} 
                                   & Weak Randomness                         \\ \cline{2-2} 
                                   & Transaction Order Dependence            \\ \hline
\end{tabular}
\end{table}

A classification of 28 vulnerabilities is proposed in \cite{Staderini_2020} and was further evolved to categorize a total of 33 vulnerabilities in \cite{Staderini2022}. Table \ref{tab:Mirko} presents an overview of the authors' classification, identifying the acronym and name of the vulnerability and an associated CWE \cite{CWECommunity2009}.

\begin{table}[h]
\centering
\caption{Classification proposed in \cite{Staderini2022}.}
\label{tab:Mirko}
\begin{tabular}{|
>{\columncolor[HTML]{FFFFFF}}l 
>{\columncolor[HTML]{FFFFFF}}l |
>{\columncolor[HTML]{FFFFFF}}l |}
\hline
\multicolumn{1}{|l|}{\cellcolor[HTML]{FFFFFF}{\color[HTML]{333333} \textit{\textbf{Acr.}}}} & {\color[HTML]{333333} \textit{\textbf{Vulnerability name}}}                                                          & {\color[HTML]{333333} \textit{\textbf{CWE-ID}}}                                                                                                  \\ \hline
\multicolumn{1}{|l|}{\cellcolor[HTML]{FFFFFF}{\color[HTML]{333333} ELT}}                    & {\color[HTML]{333333} Ether Lost in Transfer}                                                          & \cellcolor[HTML]{FFFFFF}{\color[HTML]{333333} }                                                                                                  \\ \cline{1-2}
\multicolumn{1}{|l|}{\cellcolor[HTML]{FFFFFF}{\color[HTML]{333333} RV}}                     & {\color[HTML]{333333} Requirement Violation}                                                           & \cellcolor[HTML]{FFFFFF}{\color[HTML]{333333} }                                                                                                  \\ \cline{1-2}
\multicolumn{1}{|l|}{\cellcolor[HTML]{FFFFFF}{\color[HTML]{333333} SA}}                     & {\color[HTML]{333333} Short Addresses}                                                                 & \multirow{-3}{*}{\cellcolor[HTML]{FFFFFF}{\color[HTML]{333333} \begin{tabular}[c]{@{}l@{}}CWE-20\end{tabular}}}                        \\ \hline
\multicolumn{1}{|l|}{\cellcolor[HTML]{FFFFFF}{\color[HTML]{333333} Atx}}                    & {\color[HTML]{333333} Authorization through tx. origin}                                                & \multicolumn{1}{c|}{\cellcolor[HTML]{FFFFFF}{\color[HTML]{333333} }}                                                                             \\ \cline{1-2}
\multicolumn{1}{|l|}{\cellcolor[HTML]{FFFFFF}{\color[HTML]{333333} UEW}}                    & {\color[HTML]{333333} Unprotected Ether Withdrawal}                                                    & \multicolumn{1}{c|}{\cellcolor[HTML]{FFFFFF}{\color[HTML]{333333} }}                                                                             \\ \cline{1-2}
\multicolumn{1}{|l|}{\cellcolor[HTML]{FFFFFF}{\color[HTML]{333333} Usd}}                    & {\color[HTML]{333333} Unprotected selfdestruct}                                                        & \multicolumn{1}{c|}{\cellcolor[HTML]{FFFFFF}{\color[HTML]{333333} }}                                                                             \\ \cline{1-2}
\multicolumn{1}{|l|}{\cellcolor[HTML]{FFFFFF}{\color[HTML]{333333} }}                       & \cellcolor[HTML]{FFFFFF}{\color[HTML]{333333} }                                                        & \multicolumn{1}{c|}{\cellcolor[HTML]{FFFFFF}{\color[HTML]{333333} }}                                                                             \\
\multicolumn{1}{|l|}{\multirow{-2}{*}{\cellcolor[HTML]{FFFFFF}{\color[HTML]{333333} UWSL}}} & \multirow{-2}{*}{\cellcolor[HTML]{FFFFFF}{\color[HTML]{333333} Unprotected Write to Storage Location}} & \multicolumn{1}{c|}{\cellcolor[HTML]{FFFFFF}{\color[HTML]{333333} }}                                                                             \\ \cline{1-2}
\multicolumn{1}{|l|}{\cellcolor[HTML]{FFFFFF}{\color[HTML]{333333} VEF}}                    & {\color[HTML]{333333} Visibility of Exposed Functions}                                                 & \multicolumn{1}{c|}{\multirow{-6}{*}{\cellcolor[HTML]{FFFFFF}{\color[HTML]{333333} \begin{tabular}[c]{@{}c@{}}CWE-284\end{tabular}}}} \\ \hline
\multicolumn{1}{|l|}{\cellcolor[HTML]{FFFFFF}{\color[HTML]{333333} GR}}                     & {\color[HTML]{333333} Generating Randomness}                                                           & {\color[HTML]{333333} \begin{tabular}[c]{@{}l@{}}CWE-330\end{tabular}}                                                                 \\ \hline
\multicolumn{1}{|l|}{\cellcolor[HTML]{FFFFFF}{\color[HTML]{333333} MPRA}}                   & {\color[HTML]{333333} Missing Protection against Signature}                                            & \cellcolor[HTML]{FFFFFF}{\color[HTML]{333333} }                                                                                                  \\ \cline{1-2}
\multicolumn{1}{|l|}{\cellcolor[HTML]{FFFFFF}{\color[HTML]{333333} SM}}                     & {\color[HTML]{333333} Signature Malleability}                                                          & \cellcolor[HTML]{FFFFFF}{\color[HTML]{333333} }                                                                                                  \\ \cline{1-2}
\multicolumn{1}{|l|}{\cellcolor[HTML]{FFFFFF}{\color[HTML]{333333} Ty}}                     & {\color[HTML]{333333} Type Casts}                                                                      & \multirow{-3}{*}{\cellcolor[HTML]{FFFFFF}{\color[HTML]{333333} \begin{tabular}[c]{@{}l@{}}CWE-345 \\ \end{tabular}}}                      \\ \hline
\multicolumn{1}{|l|}{\cellcolor[HTML]{FFFFFF}{\color[HTML]{333333} CPL}}                    & {\color[HTML]{333333} DoS costly Patterns and Loops}                                                   & \cellcolor[HTML]{FFFFFF}{\color[HTML]{333333} }                                                                                                  \\ \cline{1-2}
\multicolumn{1}{|l|}{\cellcolor[HTML]{FFFFFF}{\color[HTML]{333333} Gs}}                     & {\color[HTML]{333333} Gasless send}                                                                    & \multirow{-2}{*}{\cellcolor[HTML]{FFFFFF}{\color[HTML]{333333} \begin{tabular}[c]{@{}l@{}}CWE-400\end{tabular}}}                       \\ \hline
\multicolumn{1}{|l|}{\cellcolor[HTML]{FFFFFF}{\color[HTML]{333333} BU}}                     & {\color[HTML]{333333} Blockhash Usage}                                                                 & \cellcolor[HTML]{FFFFFF}{\color[HTML]{333333} }                                                                                                  \\ \cline{1-2}
\multicolumn{1}{|l|}{\cellcolor[HTML]{FFFFFF}{\color[HTML]{333333} ML}}                     & {\color[HTML]{333333} Malicious Libraries}                                                             & \cellcolor[HTML]{FFFFFF}{\color[HTML]{333333} }                                                                                                  \\ \cline{1-2}
\multicolumn{1}{|l|}{\cellcolor[HTML]{FFFFFF}{\color[HTML]{333333} SF}}                     & {\color[HTML]{333333} Secrecy Failure}                                                                 & \cellcolor[HTML]{FFFFFF}{\color[HTML]{333333} }                                                                                                  \\ \cline{1-2}
\multicolumn{1}{|l|}{\cellcolor[HTML]{FFFFFF}{\color[HTML]{333333} TD}}                     & {\color[HTML]{333333} Timestamp Dependency}                                                            & \multirow{-4}{*}{\cellcolor[HTML]{FFFFFF}{\color[HTML]{333333} \begin{tabular}[c]{@{}l@{}}CWE-668\end{tabular}}}                       \\ \hline
\multicolumn{1}{|l|}{\cellcolor[HTML]{FFFFFF}{\color[HTML]{333333} CU}}                     & {\color[HTML]{333333} Call to the Unknown}                                                             & \cellcolor[HTML]{FFFFFF}{\color[HTML]{333333} }                                                                                                  \\ \cline{1-2}
\multicolumn{1}{|l|}{\cellcolor[HTML]{FFFFFF}{\color[HTML]{333333} }}                       & \cellcolor[HTML]{FFFFFF}{\color[HTML]{333333} }                                                        & \cellcolor[HTML]{FFFFFF}{\color[HTML]{333333} }                                                                                                  \\
\multicolumn{1}{|l|}{\multirow{-2}{*}{\cellcolor[HTML]{FFFFFF}{\color[HTML]{333333} DUC}}}  & \multirow{-2}{*}{\cellcolor[HTML]{FFFFFF}{\color[HTML]{333333} Delegatecall to the Untrusted Callee}}  & \cellcolor[HTML]{FFFFFF}{\color[HTML]{333333} }                                                                                                  \\ \cline{1-2}
\multicolumn{1}{|l|}{\cellcolor[HTML]{FFFFFF}{\color[HTML]{333333} EC}}                     & {\color[HTML]{333333} DoS by External Contracts}                                                       & \multirow{-4}{*}{\cellcolor[HTML]{FFFFFF}{\color[HTML]{333333} \begin{tabular}[c]{@{}l@{}}CWE-669\end{tabular}}}                       \\ \hline
\multicolumn{1}{|l|}{\cellcolor[HTML]{FFFFFF}{\color[HTML]{333333} AP}}                     & {\color[HTML]{333333} Arithmetic Precision Order}                                                      & \cellcolor[HTML]{FFFFFF}{\color[HTML]{333333} }                                                                                                  \\ \cline{1-2}
\multicolumn{1}{|l|}{\cellcolor[HTML]{FFFFFF}{\color[HTML]{333333} IOU}}                    & {\color[HTML]{333333} Integer Overflow or Underflow}                                                   & \multirow{-2}{*}{\cellcolor[HTML]{FFFFFF}{\color[HTML]{333333} \begin{tabular}[c]{@{}l@{}}CWE-682\end{tabular}}}                      \\ \hline
\multicolumn{1}{|l|}{\cellcolor[HTML]{FFFFFF}{\color[HTML]{333333} AJ}}                     & {\color[HTML]{333333} Arbitrary Jump}                                                                  & \cellcolor[HTML]{FFFFFF}{\color[HTML]{333333} }                                                                                                  \\ \cline{1-2}
\multicolumn{1}{|l|}{\cellcolor[HTML]{FFFFFF}{\color[HTML]{333333} FE}}                     & {\color[HTML]{333333} Freezing Ether}                                                                  & \cellcolor[HTML]{FFFFFF}{\color[HTML]{333333} }                                                                                                  \\ \cline{1-2}
\multicolumn{1}{|l|}{\cellcolor[HTML]{FFFFFF}{\color[HTML]{333333} IGG}}                    & {\color[HTML]{333333} Insufficient gas griefing}                                                       & \cellcolor[HTML]{FFFFFF}{\color[HTML]{333333} }                                                                                                  \\ \cline{1-2}
\multicolumn{1}{|l|}{\cellcolor[HTML]{FFFFFF}{\color[HTML]{333333} Re}}                     & {\color[HTML]{333333} Reentrancy}                                                                      & \cellcolor[HTML]{FFFFFF}{\color[HTML]{333333} }                                                                                                  \\ \cline{1-2}
\multicolumn{1}{|l|}{\cellcolor[HTML]{FFFFFF}{\color[HTML]{333333} RLO}}                    & {\color[HTML]{333333} Right Left Override}                                                             & \cellcolor[HTML]{FFFFFF}{\color[HTML]{333333} }                                                                                                  \\ \cline{1-2}
\multicolumn{1}{|l|}{\cellcolor[HTML]{FFFFFF}{\color[HTML]{333333} TOD}}                    & {\color[HTML]{333333} Transaction Ordering Dependence}                                                 & \cellcolor[HTML]{FFFFFF}{\color[HTML]{333333} }                                                                                                  \\ \cline{1-2}
\multicolumn{1}{|l|}{\cellcolor[HTML]{FFFFFF}{\color[HTML]{333333} UEB}}                    & {\color[HTML]{333333} Unexpected Ether Balance}                                                        & \multirow{-7}{*}{\cellcolor[HTML]{FFFFFF}{\color[HTML]{333333} \begin{tabular}[c]{@{}l@{}}CWE-691\end{tabular}}}                      \\ \hline
\multicolumn{1}{|l|}{\cellcolor[HTML]{FFFFFF}{\color[HTML]{333333} ED}}                     & {\color[HTML]{333333} Exception Disorder}                                                              & \cellcolor[HTML]{FFFFFF}{\color[HTML]{333333} }                                                                                                  \\ \cline{1-2}
\multicolumn{1}{|l|}{\cellcolor[HTML]{FFFFFF}{\color[HTML]{333333} Us}}                     & {\color[HTML]{333333} Unchecked send}                                                                  & \cellcolor[HTML]{FFFFFF}{\color[HTML]{333333} }                                                                                                  \\ \cline{1-2}
\multicolumn{1}{|l|}{\cellcolor[HTML]{FFFFFF}{\color[HTML]{333333} UV}}                     & {\color[HTML]{333333} Unchecked Call Return Values}                                                    & \multirow{-3}{*}{\cellcolor[HTML]{FFFFFF}{\color[HTML]{333333} \begin{tabular}[c]{@{}l@{}}CWE-703\end{tabular}}}                      \\ \hline
\end{tabular}
\end{table}

As we can see in Table \ref{tab:Mirko}, the additional characterization by CWE is quite helpful although it is not accompanied by a blockchain-specific classification scheme, such as SWC \cite{swc} which could help in unifying knowledge. The source of information is based on a set of four references in \cite{Staderini_2020} and in five in \cite{Staderini2022}, which do not  directly map with the state of the practice (e.g., tools for vulnerability detection). Still, the process for building the classification is insightful and helpful as a way to solidify our own classification, e.g., by allowing a verification of our own mapping to CWE.

A consolidated taxonomy is presented in \cite{Rameder2022}. The authors were able to collect 54 vulnerabilities reported from different verification tools and grouped them into 10 categories. Table \ref{tab:Rameder} overviews the taxonomy created by the authors.


\begin{table}[h!]
\caption{Vulnerability classification in \cite{Rameder2022}}
\label{tab:Rameder}
\centering
\resizebox{0.62\textwidth}{!}{
\begin{tabular}{|l|l|l|}
\hline
\textbf{Group}                                                                                                 & \textbf{Code} & \textbf{Vulnerability}                                                                                                     \\ \hline
{\begin{tabular}[c]{@{}l@{}}Malicious   Environment, \\ Transactions or Input\end{tabular}}     & 1A            & Reentrancy                                                                                                                 \\ \cline{2-3} 
                                                                                                               & 1B            & Call to the   unknown                                                                                                      \\ \cline{2-3} 
                                                                                                               & 1C            & Exact balance   dependency                                                                                                 \\ \cline{2-3} 
                                                                                                               & 1D            & Improper data   validation                                                                                                 \\ \cline{2-3} 
                                                                                                               & 1E            & Vulnerable   DELEGATECALL                                                                                                  \\ \hline
{\begin{tabular}[c]{@{}l@{}}Blockchain/Environment   \\ Dependency\end{tabular}}                & 2A            & Timestamp dependency                                                                                                       \\ \cline{2-3} 
                                                                                                               & 2B            & Transaction-ordering   dependency (TOD)                                                                                    \\ \cline{2-3} 
                                                                                                               & 2C            & Bad random number   generation                                                                                             \\ \cline{2-3} 
                                                                                                               & 2D            & Leakage of   confidential information                                                                                      \\ \cline{2-3} 
                                                                                                               & 2E            & Unpredictable state   (dynamic libraries)                                                                                  \\ \cline{2-3} 
                                                                                                               & 2F            & Blockhash   dependency                                                                                                     \\ \hline
{\begin{tabular}[c]{@{}l@{}}Exception   \& \\ Error Handling Disorders\end{tabular}}            & 3A            & \begin{tabular}[c]{@{}l@{}}Unchecked low level   \\ call/send return values\end{tabular}                                   \\ \cline{2-3} 
                                                                                                               & 3B            & Unexpected throw or   revert                                                                                               \\ \cline{2-3} 
                                                                                                               & 3C            & Mishandled out-of-gas   exception                                                                                          \\ \cline{2-3} 
                                                                                                               & 3D            & Assert, require or   revert violation                                                                                      \\ \hline
{Denial   of Service}                                                                           & 4A            & Frozen Ether                                                                                                               \\ \cline{2-3} 
                                                                                                               & 4B            & Ether lost in   transfer                                                                                                   \\ \cline{2-3} 
                                                                                                               & 4C            & DoS with block gas   limit reached                                                                                         \\ \cline{2-3} 
                                                                                                               & 4D            & DoS by exception   inside loop                                                                                             \\ \cline{2-3} 
                                                                                                               & 4E            & Insufficient gas   griefing                                                                                                \\ \hline
{\begin{tabular}[c]{@{}l@{}}Resource   Consumption\\ \& Gas Issues\end{tabular}}                & 5A            & Gas costly loops                                                                                                           \\ \cline{2-3} 
                                                                                                               & 5B            & Gas costly   pattern                                                                                                       \\ \cline{2-3} 
                                                                                                               & 5C            & \begin{tabular}[c]{@{}l@{}}High gas consumption   of \\ variable data type or declaration\end{tabular}                     \\ \cline{2-3} 
                                                                                                               & 5D            & High gas consumption   function type                                                                                       \\ \cline{2-3} 
                                                                                                               & 5E            & Under-priced   opcodes                                                                                                     \\ \hline
{\begin{tabular}[c]{@{}l@{}}Authentication   \& \\ Access Control Vulnerabilities\end{tabular}} & 6A            & Authorization via   transaction origin                                                                                     \\ \cline{2-3} 
                                                                                                               & 6B            & \begin{tabular}[c]{@{}l@{}}Unauthorized   accessibility due to \\ wrong function or state variable visibility\end{tabular} \\ \cline{2-3} 
                                                                                                               & 6C            & Unprotected   self-destruction                                                                                             \\ \cline{2-3} 
                                                                                                               & 6D            & Unauthorized Ether   withdrawal                                                                                            \\ \cline{2-3} 
                                                                                                               & 6E            & Signature based   vulnerabilities                                                                                          \\ \hline
{Arithmetic   Bugs}                                                                             & 7A            & Integer over- or   underflow                                                                                               \\ \cline{2-3} 
                                                                                                               & 7B            & Integer division                                                                                                           \\ \cline{2-3} 
                                                                                                               & 7C            & Integer bugs or   arithmetic issues                                                                                        \\ \hline
{\begin{tabular}[c]{@{}l@{}}Bad   Coding and \\ Language Specifics\end{tabular}}               & 8A            & Type cast                                                                                                                  \\ \cline{2-3} 
                                                                                                               & 8B            & Coding error                                                                                                               \\ \cline{2-3} 
                                                                                                               & 8C            & Bad coding   pattern                                                                                                       \\ \cline{2-3} 
                                                                                                               & 8D            & Deprecated source   language features                                                                                      \\ \cline{2-3} 
                                                                                                               & 8E            & Write to arbitrary   storage location                                                                                      \\ \cline{2-3} 
                                                                                                               & 8F            & Use of assembly                                                                                                            \\ \cline{2-3} 
                                                                                                               & 8G            & Incorrect inheritance   order                                                                                              \\ \cline{2-3} 
                                                                                                               & 8H            & Variable   shadowing                                                                                                       \\ \cline{2-3} 
                                                                                                               & 8I            & Misleading source   code                                                                                                   \\ \cline{2-3} 
                                                                                                               & 8J            & Missing logic,   logical errors or dead code                                                                               \\ \cline{2-3} 
                                                                                                               & 8K            & Insecure contract   upgrading                                                                                              \\ \cline{2-3} 
                                                                                                               & 8L            & \begin{tabular}[c]{@{}l@{}}Inadequate or   incorrect logging \\ or documentation\end{tabular}                              \\ \hline
{\begin{tabular}[c]{@{}l@{}}Environment   \\ Configuration Issues\end{tabular}}                 & 9A            & Short address                                                                                                              \\ \cline{2-3} 
                                                                                                               & 9B            & Outdated compiler   version                                                                                                \\ \cline{2-3} 
                                                                                                               & 9C            & Floating or no   pragma                                                                                                    \\ \cline{2-3} 
                                                                                                               & 9D            & Token API   violation                                                                                                      \\ \cline{2-3} 
                                                                                                               & 9E            & Ethereum update   incompatibility                                                                                          \\ \cline{2-3} 
                                                                                                               & 9F            & Configuration   error                                                                                                      \\ \hline
{\begin{tabular}[c]{@{}l@{}}Eliminated/Deprecated\\  Vulnerabilities\end{tabular}}              & 10A           & Callstack depth   limit                                                                                                    \\ \cline{2-3} 
                                                                                                               & 10B           & Uninitialized storage   pointer                                                                                            \\ \cline{2-3} 
                                                                                                               & 10C           & Erroneous constructor   name                                                                                               \\ \hline
\end{tabular}
}
\end{table}

 This classification is more fine-grained than the previously discussed ones. However, there are a few issues with some names given to the vulnerabilities. For instance, it is not obvious to what extent \textit{integer bugs or arithmetic issues - 7C} is different from \textit{Integer over- or underflow - 7A} or \textit{Integer division - 7B}, and there are names like \textit{Gas costly loops} and \textit{Gas costly pattern} which seem very similar. Also, names like \textit{configuration error} are quite generic and could lead to a more specific vulnerability like \textit{Environment Configuration Issues}. 
Regarding the structure itself, the taxonomy has a flat organization in which the categories do not really represent aspects at the same abstraction or conceptual level. For instance, \textit{Denial of Service} is generally considered as a type of attack or the effect of an exploited vulnerability, or \textit{Configuration Issues} is quite generic, and it does not characterize the vulnerability sufficiently. We can observe a similar issue between the names given to the categories and to the specific vulnerabilities, e.g., \textit{Bad Coding Group} versus \textit{Coding error Vulnerability}. Although the classification lists 54 vulnerabilities, it would benefit of a way for evolving and including more recent ones (e.g., via an open repository).

\subsection{Community-Based Classification Schemes}

This section discusses taxonomies or classification initiatives maintained by communities. One of the most popular ones is \textit{Smart Contract Weakness Classification} SWC \cite{swc}, a vulnerability classification scheme for smart contracts whose main goals are: i) Provide a straightforward way to classify 'weaknesses' of a smart contract; ii) Identify weaknesses that lead to vulnerabilities; iii) Define a common language to describe weaknesses in the architecture, design, and coding of smart contracts; and finally, iv) Being a way to improve the effectiveness of smart contract security analysis tools \cite{epi1470}.

In SWC, each software defect has an external relationship with another taxonomy (i.e., CWE \cite{CWECommunity2009}), and there are examples (i.e., faulty and non-faulty code) to illustrate the vulnerability and a correction. SWC is a flat list structure, where the distinction between vulnerabilities and other types of defects is many times unclear. Also, it is worthwhile mentioning that there are cases where it is difficult to distinguish whether the problem is related to the blockchain platform or to the smart contract itself (e.g., \textit{Weak Sources of Randomness from Chain Attributes}, \textit{Unencrypted Private Data On-Chain}). A positive aspect is that SWC is associated withan open repository, although, at the time of writing, the last update was made in 2018. Considering the changes and new knowledge about smart contract vulnerabilities, this means that practitioners' involvement is now impaired. For instance, the classification presented in \cite{Rameder2022} identifies several new defects that are not present in SWC.

The NCC Group initiated the Decentralized Application Security Project (DASP) in 2018, which includes a vulnerability classification scheme for smart contracts. The idea of the classification is to present the top 10 threats to smart contract security, for which a single iteration was carried precisely in 2018. Thus, it does not really reflect the whole landscape of vulnerabilities. DASP provides a short description for each class of vulnerabilities, which is accompanied by pseudo-code as a way of explaining the defects. The classification emphasizes the impact the vulnerability had in real-world scenarios (e.g., reentrancy loss estimated at 3.5M ETH ~50M USD at the time). References to real-world attacks are provided (i.e., reports, magazines, etc.), which present a historical view of vulnerability exploitation. The nomenclature is clear, although some parts of the structure are questionable. For instance, the \textit{Denial of Service} category in DASP refers to \textit{gas limit reached}, \textit{unexpected throw}, \textit{unexpected kill}, and \textit{access control breached}. The description is sometimes so short that may become ambiguous (e.g., \textit{access control breached} may refer to a vulnerability that would simply fit in \textit{Access Control}, which is another DASP category). In \cite{durieux_empirical_2020}, the authors used DASP but concluded that the categories were not sufficient to cover the vulnerabilities found.

SIGP \cite{SIGP2018} is a vulnerability classification scheme for smart contracts written in Solidity that forms the basis of 
of the work in \cite{antonopoulos_mastering_2018}. The classification considers three main elements: vulnerability, preventive technique, and a real-world example. The first element conceptually describes the reported vulnerability. It also presents the vulnerable code and explains how the attack is performed. The second element presents a solution for the problem, and the last element discusses a real-world attack in which the vulnerability was exploited. The clarity of the names used for the vulnerabilities could be improved (e.g., \textit{entropy illusion} and \textit{constructors with care} are ambiguous). There is an open repository associated, but not receiving any updated, at the time of writing. As in previous cases, there are only 16 vulnerabilities listed, which is currently far from the state of the practice.

The SMARTDEC classification \cite{SmartDec2018} originated from the experience gathered from the creation of Smartcheck \cite{tikhomirov_smartcheck_2018}. The vulnerabilities are organized into three main categories: Blockchain (i.e., vulnerabilities from  the blockchain system), Language (i.e., programming language defects), and Model (i.e., vulnerabilities caused by mistakes in the model). Each group has several entries (up to a total of 11), where each entry corresponds to a set of related vulnerabilities. The entry names are unique, although they are also quite generic, and therefore less descriptive (e.g., \textit{Trust}). The authors provide a mapping between their taxonomy and other classifications, namely DASP \cite{nccgroup_decentralized_2021}, SWC \cite{swc}, and SIGP \cite{SIGP2018}. As an example, the \textit{Arithmetic} category is related to \textit{Over/underflow} in SWC-101, DASP-3, and SP-2 and to \textit{Precision issues} in SP-15. The repository is open to contributions, although, at the time of writing, there has been no update since 2018.

\subsection{Classification Schemes used in Vulnerability Detection Research}

Research in smart contract vulnerability detection for smart contracts is generally accompanied by custom vulnerability classification schemes \cite{luu_making_2016,kalra_zeus_2018,wang_vultron_2019,ghaleb_how_2020,SMARTIAN_2021,SAILFISH_2022}. This is primarily due to lacking an appropriate and up-to-date classification standard or taxonomy. As a result, biased and limited classifications emerged, which are coupled to the context in which they were created. The next paragraphs describe the classification schemes of selected research, namely of three of the most cited vulnerability detection research works (at the time of writing and according to Google Scholar). In all of these cases, the heterogeneity is clear, as well as the divergence with other classification schemes, such as the ones previously presented in this section.

A symbolic execution tool named Oyente is proposed in \cite{luu_making_2016} with the goal of allowing practitioners to detect security vulnerabilities. Within the tool proposal the authors identify a small set of security vulnerabilities, as illustrated in Table \ref{tab:Oyente}. 

\begin{table}[h]
\centering
\caption{List of vulnerabilities in \cite{luu_making_2016}}
\label{tab:Oyente}
\begin{tabular}{|l|}
\hline
\textbf{Vulnerability name}                         \\ \hline
Mishandled   Exceptions               \\ \hline
Reentrancy                            \\ \hline
Timestamp   dependence                \\ \hline
Transaction-Ordering   Dependence     \\ \hline
\end{tabular}
\end{table}

Although the work in Oyente targets a specific set of vulnerabilities, the absence of a standard way for categorizing and naming the vulnerabilities impairs the assessment and comparison of results with other tools or approaches.

Securify \cite{tsankov_securify_2018} is a vulnerability detection tool based on symbolic execution methods, which, at the time of writing, is able to detect 37 security defects \cite{Tsankov2018}, which the tool groups by severity, as we can see in Table \ref{tab:Securify}.

\begin{table}[h]
\centering
\caption{Vulnerability classification in \cite{tsankov_securify_2018} and extended in \cite{Tsankov2018}}
\label{tab:Securify}
\begin{tabular}{|l|l|}
\hline
\textbf{Severity}         & \textbf{Vulnerability}          \\ \hline
\multirow{4}{*}{Critical} & TODAmount                       \\ \cline{2-2} 
                          & TODReceiver                     \\ \cline{2-2} 
                          & TODTransfer                     \\ \cline{2-2} 
                          & UnrestrictedWrite               \\ \hline
\multirow{7}{*}{High}     & RightToLeftOverride             \\ \cline{2-2} 
                          & ShadowedStateVariable           \\ \cline{2-2} 
                          & UnrestrictedSelfdestruct        \\ \cline{2-2} 
                          & UninitializedStateVariable      \\ \cline{2-2} 
                          & UninitializedStorage            \\ \cline{2-2} 
                          & UnrestrictedDelegateCall        \\ \cline{2-2} 
                          & DAO                             \\ \hline
\multirow{10}{*}{Medium}  & ERC20Interface                  \\ \cline{2-2} 
                          & ERC721Interface                 \\ \cline{2-2} 
                          & IncorrectEquality               \\ \cline{2-2} 
                          & LockedEther                     \\ \cline{2-2} 
                          & ReentrancyNoETH                 \\ \cline{2-2} 
                          & TxOrigin                        \\ \cline{2-2} 
                          & UnhandledException              \\ \cline{2-2} 
                          & UnrestrictedEtherFlow           \\ \cline{2-2} 
                          & UninitializedLocal              \\ \cline{2-2} 
                          & UnusedReturn                    \\ \hline
\multirow{6}{*}{Low}      & ShadowedBuiltin                 \\ \cline{2-2} 
                          & ShadowedLocalVariable           \\ \cline{2-2} 
                          & CallToDefaultConstructor?       \\ \cline{2-2} 
                          & CallInLoop                      \\ \cline{2-2} 
                          & ReentrancyBenign                \\ \cline{2-2} 
                          & Timestamp                       \\ \hline
\multirow{10}{*}{Info}    & AssemblyUsage                   \\ \cline{2-2} 
                          & ERC20Indexed                    \\ \cline{2-2} 
                          & LowLevelCalls                   \\ \cline{2-2} 
                          & NamingConvention                \\ \cline{2-2} 
                          & SolcVersion                     \\ \cline{2-2} 
                          & UnusedStateVariable             \\ \cline{2-2} 
                          & TooManyDigits                   \\ \cline{2-2} 
                          & ConstableStates                 \\ \cline{2-2} 
                          & ExternalFunctions               \\ \cline{2-2} 
                          & StateVariablesDefaultVisibility \\ \hline
\end{tabular}
\end{table}

Again, as with the previous tool, the groups and the names or vulnerability definition are non-standard, although there is an effort in to classify most of them according to SWC \cite{swc}.

Zeus is a tool based on abstract interpretation and symbolic execution \cite{kalra_zeus_2018}. Table \ref{tab:Zeus} shows the vulnerability classification performed by the authors and targeted by the tools.

\begin{table}[h]
\centering
\caption{Vulnerability classification in \cite{kalra_zeus_2018}}
\label{tab:Zeus}
\begin{tabular}{|ll|}
\hline                                                 
\multicolumn{1}{|l|}{\textbf{Group}}                       & \textbf{Vulnerability}       \\ \hline
\multicolumn{1}{|l|}{\multirow{5}{*}{Incorrect Contracts}} & Reentrancy                   \\ \cline{2-2} 
\multicolumn{1}{|l|}{}                                     & Unchecked send               \\ \cline{2-2} 
\multicolumn{1}{|l|}{}                                     & Failed send                  \\ \cline{2-2} 
\multicolumn{1}{|l|}{}                                     & Integer overflow/underflow   \\ \cline{2-2} 
\multicolumn{1}{|l|}{}                                     & Transaction state dependence \\ \hline
\multicolumn{1}{|l|}{\multirow{3}{*}{Unfair Contracts}}    & Absense of Logic             \\ \cline{2-2} 
\multicolumn{1}{|l|}{}                                     & Incorrect Logic              \\ \cline{2-2} 
\multicolumn{1}{|l|}{}                                     & Logically Correct But Unfair \\ \hline
\multicolumn{1}{|l|}{\multirow{2}{*}{Miner’s Influence}}   & Block state   dependence     \\ \cline{2-2} 
\multicolumn{1}{|l|}{}                                     & Transaction order dependence \\ \hline
\end{tabular}
\end{table}

As we can see in Table \ref{tab:Zeus}, the authors created several groups (e.g., incorrect contracts, unfair contracts), in which several defects are placed. Although this is obviously a partial classification of known vulnerabilities, the heterogeneity of the naming and definitions and also general classification structures is clear (when compared to other works), which again emphasizes the need for a more standard way of categorizing defects.

\subsection{Limitations of Current Classification Schemes}

In this section, we highlight the main gaps and limitations identified during the analysis of the different vulnerability classifications previously described, as follows: 

\begin{enumerate}[-]

\item Classifications proposed in the literature tend to have simple structures, most of them simply grouping the vulnerabilities into related groups. Many times, no groups at all are used. Such structures are often ad-hoc and consequently short-lived, resulting in limited adoption. The classifications that collect more vulnerabilities are found in \cite{Rameder2022, swc, Tsankov2018}, with \cite{Tsankov2018} grouping vulnerabilities by criticality and with \cite{Rameder2022} using conceptual groups to fit related vulnerabilities.

\item  There is a large diversity of names being used in state of the art to refer to the same vulnerability (e.g., both \textit{Integer bugs or arithmetic issues} and \textit{Integer over- or underflow} \cite{Rameder2022} refer to the same vulnerability). There are also cases in which very similar names refer to different vulnerabilities (e.g., \textit{unpredictable state} \cite{bauer_semantic_2018} refers to wrong class inheritance order defect while \textit{vulnerable state} \cite{krupp_teether_2018} refers to uninitialized storage variable defect). In some cases, the same name refers to different vulnerabilities, e.g., Transaction Order Dependency (TOD) is the name used in \cite{liao_soliaudit_2019} and in \cite{SAILFISH_2022}, which however refers respectively to "5.1.5 Transfer Amount Dependent on Transaction Order" and to "5.1.6 Transfer Recipient Dependent on Transaction Order".

\item Current classifications include several generic names that do not assist in the classification of specific defects (e.g., \textit{call to the unknown} \cite{Atzei2017} or \textit{unexpected function invocation} \cite{Chen2020a}). In several cases, unclear nomenclatures are used, such as \textit{entropy illusion}, \textit{constructors with care} \cite{SIGP2018}, or \textit{Improper Blockchain Magic Validation} \cite{Amiet2021}, which do not specify what the defect is. Another example is \textit{Style guide violation} \cite{zhang_soliditycheck_2019}, which is not even clear whether it is referring to bad practice or a vulnerability.

\item Regarding vulnerability classification, current research appears to be falling far behind the state of the practice. Current vulnerability detection tools identify several vulnerabilities (e.g., Securify2 \cite{Tsankov2018}) that do not fit in relatively well-established classifications, such as DASP \cite{DASP}, or SWC \cite{swc}.

\item Current classifications do not involve active community participation, and we observed little to no participation at all in several classifications. Thus, it is fundamental that a classification can be easily maintained and evolve to integrate new vulnerabilities or even has the possibility of structurally changing (i.e., versioning is also required). This reduced community participation is the main reason why the most popular classification initiatives, like SWC \cite{swc} or DASP \cite{DASP}, are currently far behind the detection capabilities of vulnerability detection tools. 

\item Classifications originated from vulnerability detection tools sometimes use names that are biased towards the tool's capabilities, which is fully acceptable from a tool perspective, but for broader goals (e.g., tool benchmarking), a vulnerability classification must be independent of specific tools' capabilities. For instance, Osiris \cite{torres_osiris_2018} is a tool for detecting vulnerabilities related with integer values and naturally focuses on a few types of issues affecting integer manipulation. Thus, the naming used is very specific of this context and also does not capture the larger picture (e.g., issues affecting other types of numbers may be related, but are not represented).

\item The high heterogeneity of names used across various tools, community efforts, and research initiatives creates a significant obstacle to understanding which tools perform better. Although initiatives exist to assess the effectiveness of the vulnerability detection tools, they all faced difficulties in adopting a uniform, fine-grained taxonomy for defects.

\item Many times, taxonomies mix the characteristics of a certain vulnerability with the effect of exploiting it or with how it is exploited, or its impact, and use category names like \textit{Denial of Service}, which is basically a consequence of the activation of a certain vulnerability. This is not necessarily wrong, but it may contribute to a non-uniform taxonomy and possibly error-prone from the point of view of the taxonomy's user.

\item Classification structures are often constructed with different degrees of granularity. Some structures have general categories, while others have more specific categories. This inconsistent categorization of  poses difficulties and complexities for practitioners and tool developers, as they end up creating new classifications. Overall, a broader view on vulnerability detection is needed to foster the longevity of a particular taxonomy, accompanied with the possibility of evolving it. 

\end{enumerate}

\section{OpenSCV Construction Process}
\label{sec:mechanisms}

\begin{figure}[ht!]
    \centering
    \includegraphics[scale=0.8]{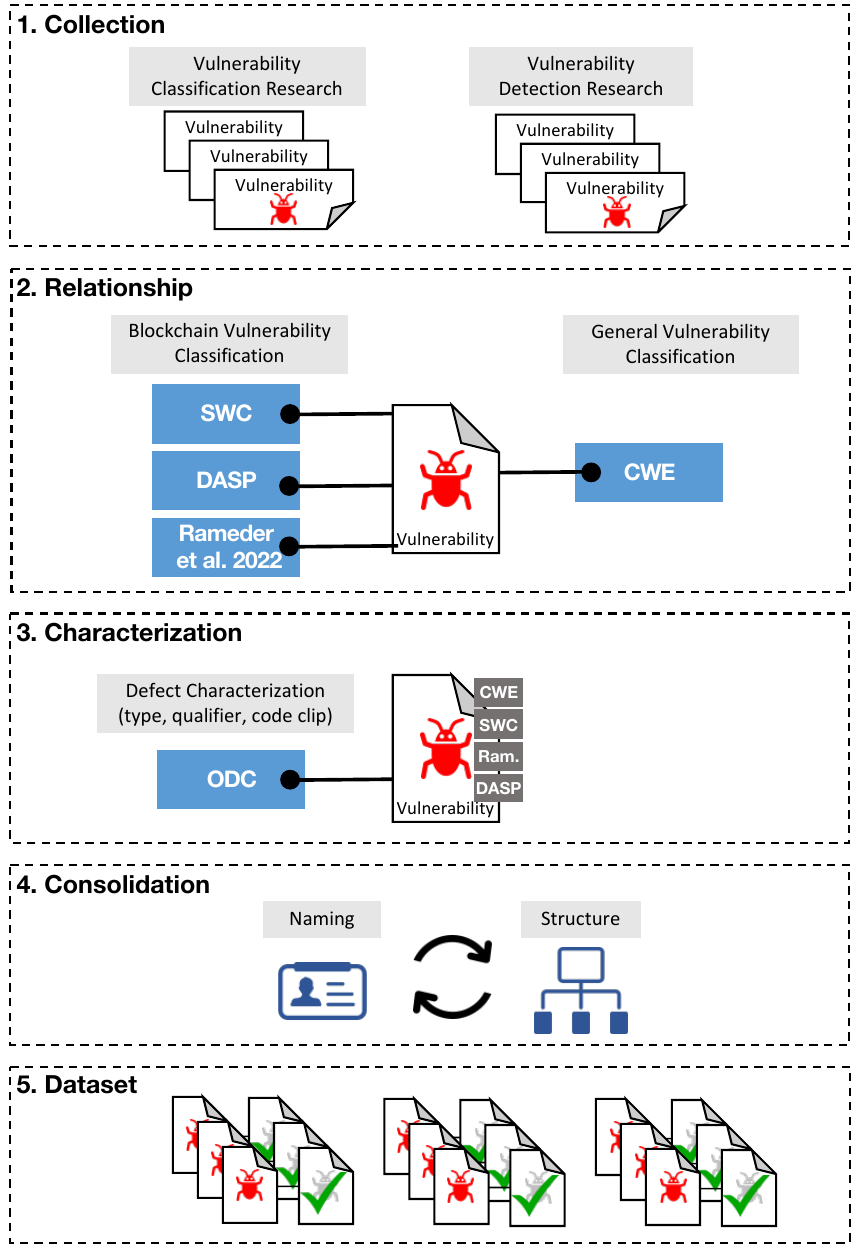}
    \caption{Taxonomy construction process.}
    \label{fig:taxonomyconstruction}
\end{figure}

This section describes the process followed to build the OpenSCV taxonomy. Overall, it was an iterative and incremental process during which we kept general taxonomy quality properties (e.g., the ones discussed in Section \ref{sec:SectionII}) in perspective, while going through all the construction phases. As mentioned in Section \ref{sec:intro}, we use the general term \textit{vulnerability} to refer to vulnerabilities and also to software defects considered in the literature to be associated with high-security risks. Figure \ref{fig:taxonomyconstruction} overviews the process, which consists of the following phases:

\begin{enumerate}[i)]
    \item Vulnerability information collection;
    \item Vulnerability relationship with other classifications;
    \item Vulnerability characterization (defect type, qualifier, and code clip example);
    \item Structural and nomenclature consolidation;
    \item Dataset construction.
\end{enumerate}

Regarding the first phase (\textbf{vulnerability information collection}), visible on the top of Figure \ref{fig:taxonomyconstruction}), the main goal was to gather an up-to-date, heterogeneous, and non-curated list of vulnerabilities that affect smart contracts. This list allowed us to understand the naming and classification heterogeneity, which is essential to build an integrated vision and ultimately reach a meaningful taxonomy. Thus, we began by using Google Scholar to try to identify research work on \textit{smart contract vulnerability classification} (e.g., taxonomies and defect classification schemes). We then proceeded to search for research targeting \textit{smart contract vulnerability detection}, which resulted in the identification of 49 research papers, which are mostly materialized in tools and that we summarize in Table \ref{tab:tools}. The identified works refer to research carried out from October 2016 to January 2023 and resulted in the collection of 357 vulnerability definitions. It is worthwhile mentioning that the identified research also led us to the identification of community-oriented initiatives, namely SWC \cite{swc} and DASP \cite{DASP}, 
which recurrently appear in the literature. There are other initiatives, such as SIGP \cite{SIGP2018} or  SMARTDEC \cite{SmartDec2018}, which seem to have less expression.

\begin{table}
\centering
\caption{Vulnerability detection tools identified in the state of the art.}
\label{tab:tools}
\resizebox{0.8\textwidth}{!}{%
\begin{tabular}{|l|l|l|l|}
\hline
\textbf{Categories}                  & \textbf{Technique}                           & \textbf{Tool Name}         & \textbf{Reference}                \\ \hline
\multirow{7}{*}{Formal Verification} & \multirow{6}{*}{Model Checking}             & EthSemantics               & \cite{bauer_semantic_2018}        \\ \cline{3-4} 
                                     &                                             & FSolidM                    & \cite{mavridou_designing_2018}    \\ \cline{3-4} 
                                     &                                             & SmartPulse                 & \cite{smartpulse_2021}            \\ \cline{3-4} 
                                     &                                             & VeriSmart                  & \cite{so_verismart_2020}          \\ \cline{3-4} 
                                     &                                             & VeriSolid                  & \cite{mavridou_verisolid_2019}    \\ \cline{3-4} 
                                     &                                             & Zeus                       & \cite{kalra_zeus_2018}            \\ \cline{2-4} 
                                     & Theorem Proving                             & ---              & \cite{ayoade_smart_2019}          \\ \hline
\multirow{5}{*}{Machine Learning}    & \multirow{5}{*}{Classical Machine Learning} & ContractWard               & \cite{wang_contractward_2020}     \\ \cline{3-4} 
                                     &                                             & ---           & \cite{momeni_machine_2019}        \\ \cline{3-4} 
                                     &                                             & SoliAudit                  & \cite{liao_soliaudit_2019}        \\ \cline{3-4} 
                                     &                                             & ---                & \cite{song_machine_2019}          \\ \cline{3-4} 
                                     &                                             & xFuzz                      & \cite{Li2023}                     \\ \hline
\multirow{28}{*}{Software Testing}   & \multirow{15}{*}{Fuzzing}                   & ContractFuzzer             & \cite{jiang_contractfuzzer_2018}  \\ \cline{3-4} 
                                     &                                             & DEPOSafe                   & \cite{Ji2020}                     \\ \cline{3-4} 
                                     &                                             & Echidna                    & \cite{grieco_echidna_2020}        \\ \cline{3-4} 
                                     &                                             & Etherolic                  & \cite{ashouri_etherolic_2020}     \\ \cline{3-4} 
                                     &                                             & EthRacer                   & \cite{kolluri_exploiting_2019}    \\ \cline{3-4} 
                                     &                                             & GasFuzzer                  & \cite{ashraf_gasfuzzer_2020}      \\ \cline{3-4} 
                                     &                                             & Reguard                    & \cite{liu_reguard_2018}           \\ \cline{3-4} 
                                     &                                             & Sereum                     & \cite{Rodler2019}                 \\ \cline{3-4} 
                                     &                                             & sFuzz                      & \cite{nguyen_sfuzz_2020}          \\ \cline{3-4} 
                                     &                                             & SMARTIAN                   & \cite{SMARTIAN_2021}              \\ \cline{3-4} 
                                     &                                             & SODA                       & \cite{Chen2020a}                  \\ \cline{3-4} 
                                     &                                             & Solanalyser                & \cite{akca_solanalyser_2019}      \\ \cline{3-4} 
                                     &                                             & SoliAudit                  & \cite{liao_soliaudit_2019}        \\ \cline{3-4} 
                                     &                                             & Vultron                    & \cite{wang_vultron_2019}          \\ \cline{3-4} 
                                     &                                             & xFuzz                      & \cite{Li2023}                     \\ \cline{2-4} 
                                     & \multirow{2}{*}{Mutation Testing}                   & Deviant                    & \cite{chapman_deviant_2019}       \\ \cline{3-4} 
                                     &                                             & Extended Univ. Mutator & \cite{andesta_testing_2020}       \\ \cline{2-4} 
                                     & \multirow{11}{*}{Symbolic Execution}        & EthRacer                   & \cite{kolluri_exploiting_2019}    \\ \cline{3-4} 
                                     &                                             & Osiris                     & \cite{torres_osiris_2018}         \\ \cline{3-4} 
                                     &                                             & Oyente                     & \cite{luu_making_2016}            \\ \cline{3-4} 
                                     &                                             & DEPOSafe                   & \cite{Ji2020}                     \\ \cline{3-4} 
                                     &                                             & Pluto                      & \cite{Ma2022}                     \\ \cline{3-4} 
                                     &                                             & RA                         & \cite{chinen_hunting_2020}        \\ \cline{3-4} 
                                     &                                             & SAILFISH                   & \cite{SAILFISH_2022}              \\ \cline{3-4} 
                                     &                                             & sCompile                   & \cite{chang_scompile_2019}        \\ \cline{3-4} 
                                     &                                             & SmartScopy                 & \cite{feng_precise_2019}          \\ \cline{3-4} 
                                     &                                             & teEther                    & \cite{krupp_teether_2018}         \\ \cline{3-4} 
                                     &                                             & Vultron                    & \cite{wang_vultron_2019}          \\ \hline
\multirow{17}{*}{Static Analysis}    & \multirow{6}{*}{Abstract Interpretation}    & MadMax                     & \cite{grech_madmax_2020}          \\ \cline{3-4} 
                                     &                                             & OpenBalthazar               & \cite{arganaraz_detection_2020}   \\ \cline{3-4} 
                                     &                                             & Securify                   & \cite{tsankov_securify_2018}      \\ \cline{3-4} 
                                     &                                             & SoliDetector               & \cite{Hu2023}                     \\ \cline{3-4} 
                                     &                                             & Vandal                     & \cite{Brent2018}                  \\ \cline{3-4} 
                                     &                                             & Zeus                       & \cite{kalra_zeus_2018}            \\ \cline{2-4} 
                                     & \multirow{4}{*}{Pattern Recognition}        & NeuCheck                   & \cite{lu_neucheck_2019}           \\ \cline{3-4} 
                                     &                                             & SmartCheck                 & \cite{tikhomirov_smartcheck_2018} \\ \cline{3-4} 
                                     &                                             & SolidityCheck              & \cite{zhang_soliditycheck_2019}   \\ \cline{3-4} 
                                     &                                             & Vrust                      & \cite{Cui2022}                    \\ \cline{2-4} 
                                     & \multirow{7}{*}{Taint Analysis}             & Clairyoyance               & \cite{ye_clairvoyance_2020}       \\ \cline{3-4} 
                                     &                                             & EasyFlow                   & \cite{gao_easyflow_2019}          \\ \cline{3-4} 
                                     &                                             & Ethainter                  & \cite{brent_ethainter_2020}       \\ \cline{3-4} 
                                     &                                             & EthPloit                   & \cite{zhang_ethploit_2020}        \\ \cline{3-4} 
                                     &                                             & Osiris                     & \cite{torres_osiris_2018}         \\ \cline{3-4} 
                                     &                                             & Sereum                     & \cite{Rodler2019}                 \\ \cline{3-4} 
                                     &                                             & Slither                    & \cite{feist_slither_2019}         \\ \hline
\end{tabular}
}
\end{table}

In the second phase, we analyzed the \textbf{vulnerability relationship with other classifications} by going through each of the identified vulnerabilities and mapping them to popular smart contract  vulnerability classification schemes, namely SWC \cite{swc} and DASP \cite{nccgroup_decentralized_2021}. We also selected, from the state of the art in vulnerability classification, what is, to the best of our knowledge, the currently largest and most recent vulnerability classification scheme proposed by \cite{Rameder2022}. Then we resorted to a broader security-related classification, namely the Common Weakness Enumeration (CWE) \cite{CWECommunity2009}, which provides us with a non-domain-specific view of each defect. Although the action consists of simply mapping vulnerabilities, it actually contributes to the characterization of each vulnerability. This may be useful for later taxonomy consolidation purposes (e.g., by merging defects that are the same but are represented with different names). Obviously, mapping the identified vulnerabilities to existing classifications also allow us to understand the exact coverage of existing classifications or disparities against the current state of the art or practice. 

The third phase – \textbf{vulnerability characterization (defect type, qualifier, and code clip example)} – has the direct goal of detailing the vulnerability according to its nature and also by example, which allows for clarity of the explanation and may help in cases where the vulnerability description and remaining attributes are inadvertently left unclear. Regarding (\textit{vulnerability nature}), we resort to the Orthogonal Defect Classification, namely to the 'defect type' attribute, which generally characterizes the type of defect and can correspond to Assignment/Initialization, Checking, Algorithm/Method, Function/Class/Object, Interface/O-O Messages, Timing/Serialization, Relationship \cite{ODC}. We also make use of the 'defect qualifier' attribute, which characterizes the state of the implementation before a correction, namely if the defect refers to missing, wrong, extraneous code. We also use the ODC extensions, as proposed in \cite{odcExtension}, for defects that relate with other aspects (e.g.,defect types related with the process followed during compilation, or management of libraries). For each identified defect, we also extracted a code excerpt (when made available by the authors) that could represent the issue, as a way to reduce or eliminate any possible ambiguities that could still be present. For the cases where no defect was made available and the description allowed to build one, we created a Solidity code example as a way of further illustrating the defect. Thus, all of the identified vulnerabilities in OpenSCV are associated with a code example.

The fourth phase \textbf{naming and structural consolidation}, consists of two steps: the attribution of names to the vulnerabilities, and; ii) their organization in a tree structure. In the first step, we merged defects that referred to the same issue (despite being named differently by different authors. This required going through the names and descriptions of the different defects and, whenever provided by the authors, also analyzing the corresponding vulnerable code to understand if it referred to the same defect or not. The additional characterization (e.g., ODC) helped in such grouping. Obviously, during this step several adjustments to the characterization of the defects were made, as well as corrections to the defects' relationships with other classifications. Figure \ref{fig:code-example} shows an example of the same vulnerability named differently by different authors. In \cite{Brent2018} named it  \textit{Unsecured Balance} (Figure \ref{fig:code-example}.a)) and it basically consists of a misnamed constructor while \cite{zhang_soliditycheck_2019} named it \textit{Missing constructor} (Figure \ref{fig:code-example}.b)), where we observe that it is actually a wrong name used during the definition of the constructor.  So, besides the names we actually see that the definitions provided may not be really accurate sometimes. In this particular case, and as an example, we named this defect as "Wrong Constructor Name". Thus, during this step, we defined an initial name for each of the defects, based on the name given by the authors of the respective paper, on the names presented in the corresponding related classifications (i.e., DASP, SWC, CWE), and on the ODC classification.


\begin{figure*}
\centering
\vspace{0.5cm}
\begin{minipage}{0.5\linewidth}
\textbf{a)}
\begin{verbatim}
contract TaxMan \{
   address private owner;
   ...
   function TaxOffice() {
       owner = msg.sender;
    }
    function collectTaxes() public {
        require(msg.sender == owner);
        owner.send(tax);}
}
\end{verbatim}
\end{minipage}\hfill
\begin{minipage}{0.5\linewidth}
\textbf{b)}
\begin{verbatim}
contract Foo 
{
   address public owner;
   ...
   function foo() public
   {
      owner = msg.sender;
      ...
   }
}
\end{verbatim}
\end{minipage}
\caption{The same vulnerability named and also described differently in: a) \cite{Brent2018}; b)  \cite{zhang_soliditycheck_2019} }
\label{fig:code-example}
\end{figure*}

\begin{figure*}[h!]
    \centering
    \includegraphics[scale=0.5]{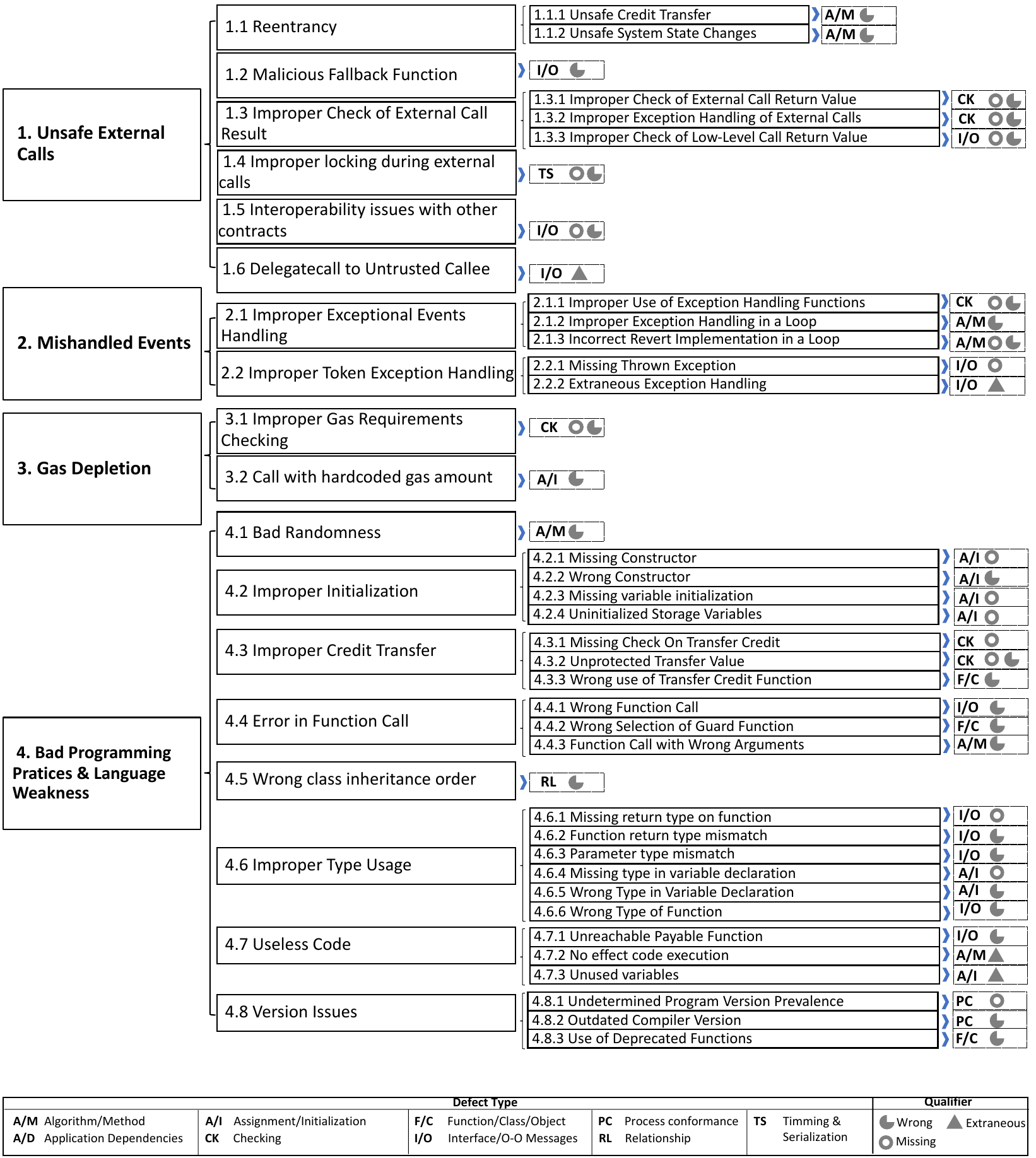}
    \caption{Taxonomy of Smart Contract Vulnerabilities (Part 1 of 2)}
    \label{fig:taxonomytree1}
\end{figure*}

\begin{figure*}[h!]
    \centering
    \includegraphics[scale=0.5]{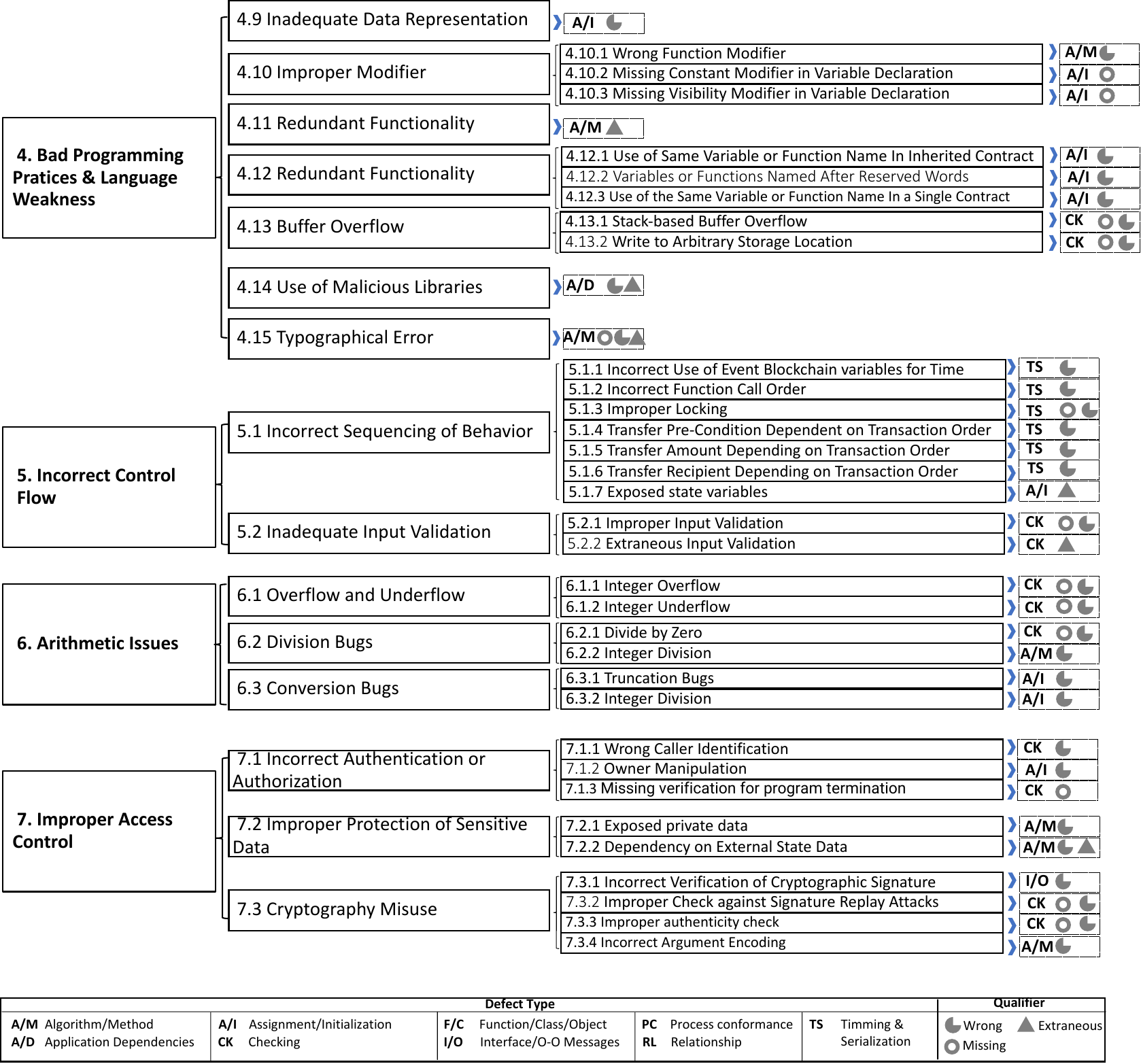}
    \caption{Taxonomy of Smart Contracts' Vulnerabilities (Part 2 of 2)}
    \label{fig:taxonomytree2}
\end{figure*}

In the second step of the fourth phase (i.e., structural consolidation), we defined a hierarchical structure for the taxonomy based on the merged vulnerabilities and preliminary naming. During this step, names were further adjusted for clarity and also to better fit in the categories being created. The final result is visible in Figure \ref{fig:taxonomytree1} and Figure \ref{fig:taxonomytree2}. As we can see in both Figures, OpenSCV consists of three levels. The first one (at the left-hand side of both Figure \ref{fig:taxonomytree1} and Figure \ref{fig:taxonomytree2}) contains the higher level categories, the intermediate one is hybrid and contains groups (i.e., subcategories) of vulnerabilities as well as a few isolated vulnerabilities. All items at the last level (at the right-hand side of Figure \ref{fig:taxonomytree1} and Figure \ref{fig:taxonomytree2}) represent vulnerabilities. Each vulnerability identified in the tree is labeled with several symbols that characterize it in terms of ODC defect type and ODC qualifier.

To build the taxonomy structure, we followed a bottom-up process and began by grouping the defects of similar nature, which allowed us to create a set of categories, such as \textit{reentrancy}, \textit{useless code}, or \textit{improper type usage}, for example. Certain defects could not really be grouped, such as \textit{Use of Malicious Libraries} or \textit{Inadequate Data Representation}, although at the same time many of them sounded like higher-level defects (i.e., siblings were expected). Thus, for the time being, we opted not to keep these vulnerabilities at the bottom layer (e.g., by creating a subcategory with a single vulnerability). After this, the same procedure was applied at this current intermediate level to reach the definition of the higher-level categories.

The whole taxonomy construction process was iterative and required the involvement of 2 Experienced Researchers and 1 Early Stage Researcher. During the process, several adjustments to names were performed for further clarity and consistency across all axis of the taxonomy.
Obviously, this is a continuous effort, which is now open to the community participation via our github repository \cite{openscvGithub}, and the current shape of the taxonomy may evolve to incorporate further vulnerabilities. It is worthwhile mentioning that, during this process, we observed that the integration of new works on vulnerability detection was a major contributor to the definition of the taxonomy, and this is the reason why we intend to be continuously integrating new works on vulnerability detection and mapping their information into new versions of our taxonomy, possibly making naming and structural adjustments as a consequence of such integration. Currently, our taxonomy lists 76 vulnerabilities and is available at \cite{openscvSite}, where all the mapped works are identified, as well as the vulnerability names used by those works. We allow for easy integration of new works, and the infrastructure is not only ready to support naming and structural changes, as well as corrections to possible errors.

The fifth phase refers to the \textbf{dataset constrution}, where we aimed at obtaining multiple real examples of smart contracts that match the defects present in our taxonomy. At the time of writing, the goal is simply to have a preliminary version of the dataset by gathering multiple real examples of contracts (i.e., a vulnerable contract and the corresponding correction) per each of the different defects present in the taxonomy. Indeed, each defect may be present in different forms (i.e., different implementations), and vulnerability detection tools may be able to detect just some of the forms. For this collection process, we directly used examples from the collected papers themselves (whenever complete contracts were made available). In some cases, SWC had usable examples also. All collected contracts present in our dataset pass through the compilation phase. Our intention is to provide an initial basis for researchers to use and, at the same time, provide the possibility of further examples (ideally, different forms of the same vulnerability) being added to the dataset.

\section{Taxonomy Levels and Defects Description}

In this section, we traverse the taxonomy tree and present a brief description of all categories and individual vulnerabilities, which are identified according to the respective numbers in Figure \ref{fig:taxonomytree1} and Figure \ref{fig:taxonomytree1}. In order to keep the used space under reasonable limits, most of the descriptions consist of brief explanations. For further information, the reader may refer to \cite{openscvSite} for further details and also examples. The exception to this is the first vulnerability discussed in the text (i.e., \textit{1.1.1 Unsafe credit transfer}), which, for illustrative purposes, we present with complete detail.


\subsection*{\hl{\textbf{1. Unsafe External Calls}}}


This category represents a set of vulnerabilities in which there is an interaction between at least two contracts. 

\subsection*{\textbf{1.1 Reentrancy}}

The first subcategory is reentrancy, in which two contracts are involved: the vulnerable contract and the malicious contract. Overall, this type of vulnerability occurs when the malicious contract, after initiating a call, is allowed to make new calls to the vulnerable contract before the initial call has been completed. Thus, unexpected state changes may occur, such as depletion of credit. We identified two main types of reentrancy vulnerabilities: one type associated with loss of credit and the other one associated with unexpected state changes. This is in line with several vulnerability detection tools, such as Securify \cite{Tsankov2018}, Slither \cite{feist_slither_2019_git}, or \cite{momeni_machine_2019} \cite{feist_slither_2019} which also distinguish these two cases (although using different names).

\subsubsection*{1.1.1 Unsafe Credit Transfer}

Known due to the DAO attack event \cite{Siegel2016}, this vulnerability allows attackers to maliciously change balance via credit transfer calls that are allowed to take place before a previous call has been completed. Let us consider the case where a smart contract maintains the balance of several addresses, allowing the retrieval of funds. A malicious contract may initiate a withdrawal operation which would lead the vulnerable contract to send funds to the malicious one before updating the balance of the malicious contract. On the malicious contract side, funds would be accepted, and a new withdrawal could be initiated (before the balance had been updated on the vulnerable contract side). As a consequence, the malicious contract could withdraw funds multiple times, with the total sum exceeding its own funds.

Using the Orthogonal Defect Classification (ODC) as a reference, this defect can be classified as being of type \textit{Algorithm} as the nature of the defect sits in the logic created by the programmer. The ODC qualifier is defined as \textit{wrong} as the error is related to incorrect logic (i.e., not \textit{missing} or \textit{extraneous} logic), related to the order of the instructions in the code.

Regarding the relationship to CWE, we classify this vulnerability (and actually the whole reentrancy group) as CWE-841, which describes a situation where \textit {"the software supports a session in which more than one behavior must be performed by an actor, but it does not properly ensure that the actor performs the behaviors in the required sequence}". This vulnerability is also known in the literature as "reentrancy" \cite{kalra_zeus_2018, mavridou_designing_2018,bauer_semantic_2018,mavridou_verisolid_2019,Brent2018,tsankov_securify_2018,arganaraz_detection_2020,ye_clairvoyance_2020,tikhomirov_smartcheck_2018,lu_neucheck_2019,liu_reguard_2018,jiang_contractfuzzer_2018,Rodler2019,liao_soliaudit_2019,Chen2020a,ashouri_etherolic_2020,ashraf_gasfuzzer_2020,nguyen_sfuzz_2020,luu_making_2016,feng_precise_2019,wang_vultron_2019,feist_slither_2019,akca_solanalyser_2019,chinen_hunting_2020,andesta_testing_2020,chapman_deviant_2019,wang_contractward_2020,song_machine_2019, Hu2023, SAILFISH_2022,smartpulse_2021,SMARTIAN_2021, Ma2022,Li2023} , "re-entrancy with balance change" \cite{momeni_machine_2019} or "SWC-107 reentrancy" \cite{swc}.

\subsubsection*{1.1.2 Unsafe System State Changes}

This vulnerability is similar in nature to \textit{v1.1.1}, with the main difference being the fact that there is no credit involved and, thus, no impact on users' funds. Due to the way the contract is coded, a call that reaches the vulnerable contract before a previous one has ended may allow an attacker to place the program in an unexpected state, leading to various effects, depending on the type of contract involved, including performance or availability issues. This vulnerability is also known in the literature as "ReentrancyNoETH" \cite{tsankov_securify_2018}, "Reentrancy" \cite{mavridou_designing_2018, mavridou_verisolid_2019}, 
 "Re-entrancy without balance change" \cite{momeni_machine_2019}, or "SWC-107 reentrancy" \cite{swc}.


\subsection*{\textbf{1.2 Malicious Fallback Function}}

Fallback functions are functions that are executed when a program receives a call to a function whose signature does not exist, i.e., either the name does not exist or the parameters do not match the parameters of any of the existing functions. For instance, an attacker could deploy a smart contract with a malicious fallback function, which could be used to drain funds or alter the system's state. By mistake, a user could invoke it and reach a state that was not expecting to reach \cite{Chen2020a}. This vulnerability is also known in the literature as "Call to the unknown" \cite{Atzei2017, arganaraz_detection_2020, chapman_deviant_2019} or "Unexpected function invocation" \cite{Chen2020a}.

\subsection*{\textbf{1.3 Improper Check of External Call Result}}

This category groups vulnerabilities that verify the execution of external contracts in an improper manner (i.e., verification is wrong or even missing), which affects the subsequent logic of the calling contract. The result of invoking a certain external operation should be verified, first of all, because it may simply fail, but especially because the called operation may be malicious (or may just have been poorly coded, resulting in an unexpected result); thus, the direct use of the result may lead to unexpected behavior.

\subsubsection*{1.3.1 Improper Check of External Call Return Value}

This defect consists of an incorrect (or missing) verification of the returned value from the external execution of a contract. When a smart contract invokes another one, the returned value should be verified because the called operation may return an unexpected value (i.e., either because the callee is malicious or may just have been poorly coded, resulting in an unexpected result) \cite{Chen2020a}. This vulnerability is also known in the literature as 
"Unchecked call return value" \cite{Zheng2021},
"Unused return" \cite{tsankov_securify_2018, momeni_machine_2019}, 
 "Unchecked external call"  \cite{tikhomirov_smartcheck_2018},  "No check after contract invocation" \cite{Chen2020a},
  "Call-depth" \cite{liao_soliaudit_2019},
 "Not checked return values" \cite{andesta_testing_2020}, ,
 "Call-stack Depth Attack" \cite{wang_contractward_2020,song_machine_2019} or "SWC-104 Unchecked Call Return Value" \cite{swc}.

\subsubsection*{1.3.2 Improper Exception Handling of External Calls}

In the case of this defect, the problem resides in the incorrect (or missing) handling of exceptional behavior thrown by a call (i.e., instead of residing in the handling of values, as in the case of vulnerability \textit{v1.3.1}). The improper verification of exceptions thrown by the callee may lead to unexpected behavior in the caller contract. There are various reasons why the callee may exhibit exceptional behavior. For instance, the callee could be under malicious control, the execution of the transaction could activate a fault in the callee contract, the transaction could be terminated due to reaching the gas limit, or the callee contract may have been terminated (e.g., after a software fault has been detected in the contract). This vulnerability is also known in the literature as  "DoS by external contract" \cite{zhang_soliditycheck_2019,tikhomirov_smartcheck_2018,lu_neucheck_2019}, "Denial of service" \cite{ashouri_etherolic_2020,andesta_testing_2020} or "SWC-113	
DoS with Failed Call" \cite{swc}.

\subsubsection*{1.3.3 Improper Check of Low-Level Call Return Value}

Languages like Solidity offer the possibility of using low-level calls that operate over raw addresses. Such calls do not verify that the code exists or the success of the calls. Thus, its use may lead to unexpected behavior \cite{large_lowlevel_2023}. As a result, using such calls can be risky and should be avoided in most cases. This vulnerability is also known in the literature as "LowLevelCalls" \cite{tsankov_securify_2018, liao_soliaudit_2019}, "Unchecked calls" \cite{feng_precise_2019, Hu2023}, "InlineAssembly" \cite{liao_soliaudit_2019}, "Usage of low-level calls" \cite{momeni_machine_2019}, or {"Check-effects"} \cite{liao_soliaudit_2019}.

\subsection*{\textbf{1.4	Improper Locking During External Calls}}

A vulnerable contract uses a lock mechanism in an erroneous manner, which may cause deadlocks. This may result, for instance, in the impossibility of executing transfers and eventually in Denial of Service \cite{mavridou_verisolid_2019}. This vulnerability is also known in the literature as "Deadlock-freedom" \cite{mavridou_verisolid_2019} or "SWC-132 Unexpected Ether balance"\cite{swc}.

\subsection*{\textbf{1.5 Interoperability Issues with Other Contracts}}

This issue relates to interoperability issues between contracts built in different language versions. Newer contracts may execute or inherit discontinued functionality present in older contracts \cite{interoperable_2021}. For instance, Solidity has introduced the operation code \texttt{STATICCALL} to allow a contract to call another contract (or itself) without modifying the state. Starting from V0.5.0, \textit{pure} and \textit{view} functions must now be called using the code \texttt{STATICCALL} instead of the usual \texttt{CALL} code. Consequently, when defining an interface for older contracts, the programmer should only use \textit{view} instead of constant in the case s/he is absolutely sure that the function will work with \texttt{STATICCALL} \cite{solidity_0_8_17}. This vulnerability is also known in the literature as "AssemblyUsage" \cite{tsankov_securify_2018, momeni_machine_2019}.

\subsection*{\textbf{1.6	Delegatecall to Untrusted Callee}}

Calling untrusted contracts using the delegate feature is generally highly problematic because it opens the possibility for the called contract to change sensitive variables (e.g., \texttt{msg.data} or \texttt{sender}) of the source contract \cite{jiang_contractfuzzer_2018}. This type of issue has been most notably known as the Parity hack, which allowed attackers to reset the ownership and usage arguments of existing user wallets  \cite{krupp_teether_2018}. This vulnerability is also known as "Unrestricted delegate call" \cite{tsankov_securify_2018}, 
"Dangerous delegate call"  \cite{jiang_contractfuzzer_2018, ashraf_gasfuzzer_2020}, "Unchecked delegate call function" \cite{nguyen_sfuzz_2020},
"Code injection"  \cite{krupp_teether_2018}, "Control-flow Hijack" \cite{SMARTIAN_2021},
"Delegated call"  \cite{andesta_testing_2020,Li2023,Hu2023}, 
"Cross Program Invocation"	\cite{Cui2022},
"Tainted delegatecall"  \cite{brent_ethainter_2020} or "SWC-112 Delegatecall to Untrusted Callee" \cite{swc}.

\subsection*{\hl{\textbf{2. Mishandled Events}}}

This category includes a set of vulnerabilities in which exceptional events are mishandled. In Solidity, there are specific functions that can be used to verify if certain conditions exist and to throw exceptions in the case the conditions are not met, namely \texttt{require} and \texttt{assert}. There are, however, fundamental differences. When the  \texttt{require} function returns false, all executed changes are reverted, and all remaining gas fees are refunded. When the \texttt{assert} function returns false, it reverts all changes but consumes all remaining gas. However, such differences have become a frequent source of problems \cite{solc-verify_2020}.

\subsection*{\textbf{2.1 Improper Exceptional Events Handling}}

This first group of vulnerabilities is directly related to exceptional events, which, when mishandled, are many times linked to the loss of atomicity in operations as well as other effects, such as excessive gas consumption or unauthorized access.

\subsubsection*{2.1.1 Improper Use of Exception Handling Functions}

Diverse runtime errors (e.g., out-of-gas error, data type overflow error, division by zero error, array-out-of-index error, etc.) may happen after a compiled smart contract is deployed. However, Solidity has many functions for error handling (e.g., \texttt{throw}, \texttt{assert}, \texttt{require}, \texttt{revert}), but their correct use relies on the experience and expertise of the developer. This defect occurs when the developer misuses the handling exception functions, which can lead the program to unexpected behavior. 
This vulnerability is also known in the literature as  
"Mishandled exceptions" \cite{bauer_semantic_2018, SMARTIAN_2021,zhang_soliditycheck_2019,  nguyen_sfuzz_2020, luu_making_2016}, "UnhandledException" \cite{ashouri_etherolic_2020,tsankov_securify_2018}, "Exception disorder \cite{jiang_contractfuzzer_2018}", or
"Exception state" \cite{momeni_machine_2019}.

\subsubsection*{2.1.2 Improper Exception Handling in a Loop}

This vulnerability occurs when a transaction is excessively large (i.e., it executes too many statements) and may lead to excessive costs. For instance, when one of the statements in a transaction fails (e.g., due to a software bug), the transaction will not be packaged into a block, and the consumed gas will not be returned to the user (and actually the concluded operations are reverted and must be executed again). Thus, such kinds of transactions should be decomposed into smaller parts so that the likelihood of success increases and the negative effects associated with the failure cases diminish. This vulnerability is also known in the literature as "CallInLoop" \cite{tsankov_securify_2018}, "Revert DOS" \cite{smartpulse_2021},
"Costly loop" \cite{zhang_soliditycheck_2019, tikhomirov_smartcheck_2018, lu_neucheck_2019},
"Multiple calls in a single transaction"  \cite{momeni_machine_2019},  "UnboundedMassOperation" \cite{grech_madmax_2020} or "SWC-128: DoS With Block Gas Limit" \cite{swc}.

\subsubsection*{2.1.3 Incorrect Revert Implementation in a Loop}

In the case of this vulnerability, the developer incorrectly specifies how the revert operation should be handled (in the context of a loop or a transaction composed of multiple operations), which ends up in a partial revert of the whole set of operations that should be reverted. This vulnerability is also known in the literature as  "Nonisolated calls (wallet griefing)" \cite{grech_madmax_2020}, "Push DOS " \cite{smartpulse_2021}, or "SWC-126 Insufficient Gas Griefing" \cite{swc}.

\subsection*{\textbf{2.2 Improper Token Exception Handling}}

The ERC-20 standard \cite{Vogelsteller2015} provides functionalities to exchange tokens. Besides describing the functionalities, the standard specifies good practices for developers to implement its features. Regarding the \texttt{transfer} function, exceptional events can become problematic if they are not handled properly.

\subsubsection*{2.2.1 Missing Thrown Exception}

Regarding the \texttt{transfer} function (i.e., functionality to transfer tokens from one account to another), the ERC-20 standard recommends to the developer throw an exception when a condition of the caller’s account balance does not have enough tokens to spend. This allows the caller to understand the reason for which the transfer is not completed and take appropriate action. 
This vulnerability is also known in the literature as "Non-standard Implementation of Tokens" \cite{Ji2020}, Missing the Transfer Event \cite{Chen2020a}.

\subsubsection*{2.2.2 Extraneous Exception Handling}

This type of defect refers to the implementation of extra actions compared to what is recommended in a certain specification. The specification does not recommend actions like the use of guard functions (e.g., require or assert) in addition to throwing an exception in the case when there is no balance in the caller. The extra actions might be arbitrary and incompatible with the purpose of a transfer functionality (e.g., returning true or false to report the success of the execution). This vulnerability is also known
in the literature as "Token API violation"  \cite{zhang_soliditycheck_2019, tikhomirov_smartcheck_2018}

\subsection*{\hl{\textbf{3. Gas Depletion}}}

This category groups defects that, in different ways, lead to gas depletion of the account used for the smart contract execution.

\subsection*{\textbf{3.1 Improper Gas Requirements Checking}}

This defect represents missing or wrong checking of the prerequisites (i.e., in terms of gas) for executing a certain operation, causing unnecessary processing and use of memory resources. For cost management reasons, languages offer programmers several ways to deal with the cost of the executing a certain operation in a contract. For instance, for transferring credits, Solidity provides the functions \texttt{transfer()} and \texttt{send()}, which have a limit of 2300 gas units for each execution. An alternative is to build a custom transfer function, where the gas limit is defined by a variable (e.g., \texttt{address.call.value(ethAmount).gas(gasAmount)()}). Despite having several ways of managing the program costs, it is challenging for programmers to predict which part of the code may fail. If an out-of-gas exception is triggered, the result may be unexpected behavior. This vulnerability is also known in the literature as "Send without Gas" \cite{arganaraz_detection_2020},
"Gassless send"  \cite{jiang_contractfuzzer_2018,ashraf_gasfuzzer_2020,nguyen_sfuzz_2020,feng_precise_2019,wang_vultron_2019,chang_scompile_2019,chapman_deviant_2019}, "Gas Dos" \cite{smartpulse_2021},
"Out of gas" \cite{akca_solanalyser_2019} or "SWC-126 Insufficient Gas Griefing" \cite{swc}.

\subsection*{\textbf{3.2 Call with Hardcoded Gas Amount}}

This defect refers to the impossibility of adjusting the amount of gas used by a certain program after being deployed. This issue is related to the observation that certain transfer credit in real contracts was being deployed using a fixed amount of gas (i.e., 2300 gas). If the gas cost of EVM instructions changes during, for instance, a hard fork, previously deployed smart contracts will easily break. This vulnerability is also known
in the literature as "SWC-134 Message call with hardcoded gas amount" \cite{swc}.

\subsection*{\hl{\textbf{4. Bad Programming Practices and Language Weaknesses}}}

This category represents issues that are mostly related to bad programming practices (i.e., error-prone or insecure coding practices) and language weaknesses, which are mostly related to insufficient protection mechanisms offered by the language, allowing the developers to make mistakes that could be avoided, e.g., by language constructs.

\subsection*{\textbf{4.1 Bad Randomness}}

This vulnerability is related to the use of the variables that control the blocks in a blockchain as a way of generating randomness, which is not secure. Such variables may be manipulated by miners so that the randomness is subverted, compromising the security of the blockchain, with its information becoming vulnerable to attacks. In fact, generating a strong enough source of randomness can be very challenging. The use of variables like \texttt{block.timestamp}, \texttt{blockhash}, \texttt{block.difficulty}, and other fields is problematic as these can be manipulated by miners. For example, a miner could select a specific timestamp within a delimited range, or use powerful hardware to mine several blocks quickly, choose the block that would provide an interesting hash, and drop the remaining. This vulnerability is also known
in the literature as 
"Generating Randomness" \cite{bauer_semantic_2018,tsankov_securify_2018},  
"Random generation" \cite{arganaraz_detection_2020}, "Bump Seeds"	\cite{Cui2022},
"Dependence on predictable variables"  \cite{lu_neucheck_2019},
"Bad randomness"  \cite{ashouri_etherolic_2020,Hu2023}, 
"Bad random" \cite{feng_precise_2019} or "SWC-120 Weak Sources of Randomness from Chain Attributes" \cite{swc}.

\subsection*{\textbf{4.2 Improper Initialization}}

The smart contract has resources that are either not initialized or are initialized in an incorrect manner, leading to unexpected behavior.

\subsubsection*{4.2.1 Missing Constructor}

A smart contract constructor is a function that is executed exactly once during the lifetime of a contract. It executes at deployment time and initializes state variables, performs a few necessary tasks that the specific contract requires, and sets the contract owner. If there is no constructor, the developer will have to implement such tasks manually, which is prone to security issues (e.g., variables may be set with incorrect values or forgotten, which may result in security problems). This vulnerability is also known in the literature as 
"Unsecure balance" \cite{Brent2018},  
"Missing constructor" \cite{zhang_soliditycheck_2019} or "SWC-118 Incorrect Constructor Name" \cite{swc}.

\subsubsection*{4.2.2 Wrong Constructor Name}

Contract published without a constructor because the programmer created a function, imagining that it would behave like a constructor. Usually, the construction function has sensitive code (e.g., assignment of the owner of the contract), and by declaring a wrong function name, any user can call the function, thus, causing serious security risks. This vulnerability is also known in the literature as "Constructor name"  \cite{andesta_testing_2020}, "Erroneous Constructor Name"	\cite{Hu2023} or "SWC-118 Incorrect Constructor Name" \cite{swc}.

\subsubsection*{4.2.3 Missing Variable Initialization}

This defect refers to the lack of initialization of variables that are used throughout the contract. Obviously, the effects can largely vary, depending on the variable itself and on the context in which is being used. This vulnerability is also known in the literature as 
"UninitializedStateVariable" \cite{tsankov_securify_2018}, 
"Uninitialized-local" \cite{tsankov_securify_2018} or 
"Uninitialized variables"  \cite{feist_slither_2019}.

\subsubsection*{4.2.4 Uninitialized Storage Variables}

In Solidity, state variables are assigned to memory or storage. When a state variable is declared, it is assigned to a certain storage slot. If that variable is not initialized, it will be stored in slot 0 (the first one) of the contract's storage. Thus, it may conflict with the next variable that is declared in the same slot, causing an address conflict. This latter variable will overwrite the first, leading to unexpected behavior. This is the reason why it is important to initialize all state variables in a smart contract so that they are set into the correct storage slots (and possible conflicts are avoided)\cite{antonopoulos_mastering_2018}. This vulnerability is also known
in the literature as 
"Uninitialized storage pointers" \cite{antonopoulos_mastering_2018}, "UninitializedStorage" \cite{tsankov_securify_2018,Hu2023} or "SWC-109 Uninitialized Storage Pointer" \cite{swc}.

\subsection*{\textbf{4.3 Improper Credit Transfer}}

This category groups defects which are generally related to improper credit transfer operations.

\subsubsection*{4.3.1 Missing Check On Transfer Credit}

This defect refers to the absence of verification after a transfer event, which can lead to an erroneous vision of the correct balance of the account. Indeed, the balance of the account may not reflect the currency transferred in an exact manner, leading to potential errors and opening the door to security issues. This vulnerability is also known in the literature as 
"Unchecked send" \cite{kalra_zeus_2018,smartpulse_2021,Brent2018,akca_solanalyser_2019,lu_neucheck_2019}.

\subsubsection*{4.3.2 Unprotected Transfer Value}

The \texttt{transfer} function uses a numeric variable for transfers and may be vulnerable if it does not protect or specify limits for the values. When attribute \texttt{address.balance} is used for identifying the amount to be transferred, it will result in transferring the total balance at once, which is a high-risk operation for the cases where the amount is high \cite{zhang_ethploit_2020}. This vulnerability is also known in the literature as 
"Unchecked transfer value" \cite{zhang_ethploit_2020} ,
"Transfer forwards all gas"  \cite{tikhomirov_smartcheck_2018},  
"UnrestrictedEtherFlow" \cite{tsankov_securify_2018},
"Ether Leak" \cite{SMARTIAN_2021},
"Manipulated Balance" \cite{Hu2023},
"Multiple Send"	\cite{SMARTIAN_2021}
or 
"SWC-105 Unprotected Ether Withdrawal" \cite{swc}.

\subsubsection*{4.3.3 Wrong use of Transfer Credit Function}

Depending on the programming language, there are different ways to carry out credit transfer operations. In Solidity, \texttt{transfer} and \texttt{send} will both allow executing a credit transfer. However, in the case of a problem, \texttt{transfer}  will abort the process with an exception, whereas \texttt{send} function will return \texttt{false}, and transaction execution is continued. An attacker may manipulate the \texttt{send} function and be able to continue executing a credit transfer operation without proper authorization. This vulnerability is also known
in the literature as 
"Failed Send"  \cite{kalra_zeus_2018}, "Use of send instead of transfer"\cite{arganaraz_detection_2020}, or
"Send instead of transfer" \cite{tikhomirov_smartcheck_2018}. 



\subsection*{\textbf{4.4 Error in Function Call}}

In a blockchain context, each function in a smart contract is identified by its name, input parameters, and output parameters. Thus, these items compose the function \textit{signature}, which is used by the contracts to verify that the right function is being called. 
This category groups defects in which a developer uses a function in the wrong manner: either a wrong signature is used, wrong arguments are used, or a wrong function is called. 

\subsubsection*{4.4.1 Wrong Function Call}

The issue occurs when a contract executes a certain function at a wrong address, i.e., at the address used by another function, which, however, has the same signature as the intended function. This vulnerability is also known in the literature as "Type casts" \cite{Atzei2017}.

\subsubsection*{4.4.2 Wrong Selection of Guard Function}

\texttt{Assert} is a Solidity function, which is recommended to be used only in the development phase. Intentionally, the programmer inserts the function at a specific point in the program where a bug is suspected. If running the program results in gas depletion, the suspicion is confirmed.

Thus, this defect refers to the cases in which the \texttt{assert} function is implemented with the wrong purpose, not having the expected effect. In more severe cases, in which the programmer forgets to remove it from the code or does not replace it with \texttt{require}, the impact of this defect can be serious. This vulnerability is
also known in the literature as 
"AssertFail" \cite{liao_soliaudit_2019}, "Assertion Failure" \cite{SMARTIAN_2021} or
"SWC-110 Assert Violation" \cite{swc}.

\subsubsection*{4.4.3 Function Call with Wrong Arguments}

This defect refers to the presence of certain control characters within the arguments of a function call, namely the \textit{right-to-left override control character}, which can cause the function to execute with arguments in reverse order. This is a known issue also in other computing areas \cite{Yosifova2021}. This vulnerability is also known
in the literature as "RightToLeftOverride" \cite{tsankov_securify_2018} or "SWC-130 Right-To-Left-Override control character (U+202E)".

\subsection*{\textbf{4.5 Wrong Class Inheritance Order}}

Contracts may have inheritance relationships with other contracts. In the case of solidity, the code of the inherited contract is always executed first, e.g., so that state variables are initialized properly. Solidity uses an algorithm named C3 linearization to determine the order in which the contracts are to be executed. Developers specify the inheritance relationships in a \texttt{inherit} statement and may believe that the order in which the inherited contracts are specified in that statement reflects the order in which the linearization algorithm should work. This opens space for security issues due to the wrong order of the contract in the \texttt{inherit} statement. This vulnerability is also known
in the literature as "Unpredictable state" \cite{bauer_semantic_2018,arganaraz_detection_2020} or "SWC-125 Incorrect Inheritance Order"\cite{swc}.

\subsection*{\textbf{4.6 Improper Type Usage}}

This category groups vulnerabilities in which there is some misuse of types of data structures or functions.

\subsubsection*{4.6.1 Missing return type on Function}

This vulnerability refers to a missing return type in the definition of a smart contract interface. At runtime, if a contract that implements that interface contains two functions with the same name and arguments but have different return types, there is a chance that the wrong function will be called. This may lead to unexpected results once the calling contract receives the wrong data type \cite{zhang_soliditycheck_2019}. This vulnerability is also known in the literature as 
"ERC20Interface"\cite{tsankov_securify_2018}, 
"Unsafe inherit from token"  \cite{lu_neucheck_2019}, "Missing Return Statement"	\cite{Hu2023} or
"Incorrect ERC20 interface" \cite{momeni_machine_2019}.

\subsubsection*{4.6.2 Function Return Type Mismatch}

In this case, the developer implemented a function (starting from an interface), but it selected the wrong data type for the value to be returned (i.e., it differs from what is specified in the interface). This vulnerability is known in the literature in the context of non-fungible tokens by the name of 
"ERC721Interface" \cite{tsankov_securify_2018} or "SWC-127 Arbitrary Jump with Function Type Variable" \cite{swc}.

\subsubsection*{4.6.3 Parameter Type Mismatch}
This issue refers to a divergence regarding the types of arguments used in a function that implements an interface. In this situation, even if the call is done with the right function name and arguments, the EVM considers it to be a non-existent function error. 
This vulnerability is also known in the literature as 
"Types conversion" \cite{arganaraz_detection_2020},  
"Unindexed ERC20 event parameters" \cite{momeni_machine_2019} or  "ERC20Indexed" \cite{tsankov_securify_2018} in the context of fungible tokens.

\subsubsection*{4.6.4 Missing Type in Variable Declaration}

In Solidity, whenever a variable is declared without an associated type, the compiler infers the data type based on the assigned value. This additional computation may lead to higher costs (i.e., in gas) and memory usage and especially allows for  overflow or underflow problems to occur. For instance, the compiler may infer that a signed integer is the right datatype for a certain variable, where an unsigned integer should be used. This vulnerability is also known in the literature as 
"Unsafe type inference" \cite{zhang_soliditycheck_2019,tikhomirov_smartcheck_2018} or 
"Unsafe-type declaration" \cite{lu_neucheck_2019} .

\subsubsection*{4.6.5 Wrong Type in Variable Declaration}



This issue refers to a wrong selection of datatypes that leads to the allocation of more memory than what would be necessary for the intended function, leading to an increase in gas consumption. As an example, in Solidity, the \texttt{byte[]} type reserves 31 bytes of space for each element, whereas the \texttt{bytes} requires a single byte per element, thus being more space efficient. This vulnerability is also known in the literature as  
"byte[ ]" \cite{zhang_soliditycheck_2019},
 "Byte array" \cite{tikhomirov_smartcheck_2018}, or
"Costly bytes"  \cite{lu_neucheck_2019}.

\subsubsection*{4.6.6 Wrong Type of Function}

In Solidity, it is possible to specify a type for each function. Functions of type \texttt{view} can read data from state variables but cannot modify them, and no gas costs are involved, whereas functions of type \texttt{pure} neither can read nor modify state variables and similarly to view functions, no gas costs are associated with this type of function. This vulnerability occurs when a developer uses the wrong type for a function. For instance, there is an issue reported in Ethereum's GitHub \cite{Pure_Issue_2022} in which a state variable conversion operation (from storage to memory) inside a \texttt{pure} function results in a problem (i.e., to avoid this problem, the function type should be \texttt{view}). This vulnerability is also known in the literature as 
"Function type operators" \cite{chapman_deviant_2019}.



\subsection*{\textbf{4.7 Useless Code}}

This category groups a set of vulnerabilities in which the program contains a unit of code that, in practice, has no effect.

\subsubsection*{4.7.1 Unreachable Payable Function}

This defect refers to the case of contracts that allow the use of functions that accept credit but do not have any functionality for transacting it. They are insecure, as there is no way to recover the credit once it has been sent to the contract \cite{zhang_soliditycheck_2019}. This vulnerability is
also known in the literature as  
"Locked Ether" \cite{chapman_deviant_2019,ashouri_etherolic_2020,feist_slither_2019,tsankov_securify_2018},
"Lost ether in the transactions" \cite{arganaraz_detection_2020}, "Locked Money"  \cite{zhang_soliditycheck_2019,tikhomirov_smartcheck_2018,lu_neucheck_2019}, "Frozen Ether"	\cite{Hu2023},
"Freezing Ether" \cite{nguyen_sfuzz_2020,SMARTIAN_2021,jiang_contractfuzzer_2018,ashraf_gasfuzzer_2020},
"Be no black hole" \cite{chang_scompile_2019}, 
"Function Can/Cannot Receive Ether" \cite{chapman_deviant_2019}, "Locked Funds"	\cite{smartpulse_2021},
"Leaking Ether to Arbitraty Address"	\cite{Hu2023} or "Contracts that lock ether" \cite{momeni_machine_2019}.

\subsubsection*{4.7.2 No Effect Code Execution}

This vulnerability refers to the presence of code that has no practical purpose (i.e., it has no effect on the intended functionality). Within a smart contract, it increases the size of the program's binary code, which results in more gas consumption than would otherwise be necessary. This vulnerability is also known in the literature as 
"CallToDefaultConstructor" \cite{tsankov_securify_2018}, "Useless Assignment"	\cite{Hu2023} or "SWC-135 Code With No Effects" \cite{swc}.

\subsubsection*{4.7.3 Unused Variables}

This defect refers to the declaration of variables that are not used in the contract, which results directly in the allocation of unnecessary space in memory. As a consequence, the gas cost of executing the contract increases as well as the attack surface of the contract. Other effects are related to the readability or maintainability of the code. This vulnerability is
also known in the literature as 
"UnusedStateVariable" \cite{tsankov_securify_2018,Hu2023}, 
"Presence of unused variables" \cite{swc} or "SWC-131 Presence of unused variables" \cite{swc}.

\subsection*{\textbf{4.8 Version Issues}}

This category refers to issues that relate to the versioning of various aspects, including the use of deprecated versions of functions.

\subsubsection*{4.8.1 Undetermined Program Version Prevalence}

This defect refers to the case where the developer allows a certain contract to be compiled across multiple versions. This allows the known faults in older versions to be easily activated. \cite{zhang_soliditycheck_2019}. This vulnerability is
also known in the literature as  
"SolcVersion" \cite{tsankov_securify_2018},
"Compiler version not fixed" \cite{tikhomirov_smartcheck_2018},
"Unfixed compiler version" \cite{lu_neucheck_2019}, 
"Usage of complex pragma statement" \cite{momeni_machine_2019} or  "SWC-103 Floating Pragma" \cite{swc}.

\subsubsection*{4.8.2 Outdated Compiler Version}

Contracts that have been developed against an outdated compiler version can bring in several risks, mainly because newer versions may have resolved certain bugs or even introduced security mechanisms to avoid particular issues (e.g., the \texttt{throw} function has been disallowed in Solidity 0.5.0 and superior versions, in favor of \texttt{assert}, \texttt{require⁄}, and \texttt{revert}). This vulnerability is
also known in the literature as 
"Compiler version problem" \cite{zhang_soliditycheck_2019} ,  
"Unfixed compiler version" \cite{lu_neucheck_2019} or "SWC-102 Outdated Compiler Version" \cite{swc}.

\subsubsection*{4.8.3 Use of Deprecated Functions}

Deprecated functions are not recommended due to the fact that they are usually replaced by functions that solve known security issues or even operate in a more efficient manner (i.e., may consume less gas). As an example, \texttt{sha3} was marked as a deprecated function in Solidity 0.5 and replaced by \texttt{keccak256}, which is more secure and efficient. This vulnerability is
also known in the literature as 
"SWC-111 Use of deprecated solidity functions" \cite{swc}.





\subsection*{\textbf{4.9 Inadequate Data Representation}}

The numbers to represent the credits (e.g., Ether) can be very large (i.e., literals with many digits are difficult to read and review). Thus it is recommended that the programmer use the native resources of the language to make this representation (e.g., Solidity 10000000000000000000 for 1 ether). This vulnerability is also known in the literature as "too-many-digits" \cite{tsankov_securify_2018}.

\subsection*{\textbf{4.10 Improper Modifier}}

This group gathers defects that relate to the use of modifiers in functions and variables.

\subsection*{4.10.1 Wrong Function Modifier}

This defect refers to the case of functions that are written solely to be used by other contracts (i.e., not within the contract). Such functions should be marked with the \texttt{external} modifier instead of \texttt{public}. The \texttt{public} modifier allows both external and internal calls. Marking a function with \texttt{external} results in gas savings, as every invocation will be using \texttt{calldata} (a special memory region to store arguments, which cannot be later modified by the function) and can avoid unnecessary read and write operations to memory, which occur with internal calls (i.e., that do not use \texttt{calldata}). This vulnerability is also known in the literature as 
"external-function" \cite{tsankov_securify_2018} or "SWC-100: Function Default Visibility" \cite{swc}.

\subsection*{4.10.2 Missing Constant Modifier in Variable Declaration}


Variables that are not modified during the execution flow should be declared as constants to save gas. In the absence of the \texttt{constant} modifier, it is assumed that the variable's value can be changed. This vulnerability is also known in the literature as 
"ConstableStates" \cite{tsankov_securify_2018} or
"State  variables that could be declared as constant" \cite{momeni_machine_2019}.

\subsubsection*{4.10.3	Missing Visibility Modifier in Variable Declaration}

Variables have different visibility states, which determine the context for accessing them. In Solidity, by default, the visibility of state variables and functions is \texttt{internal}, which allows access from functions in the same contract or derived contracts. A developer that is unaware of this may create a contract that allows exposure of sensitive data or allow unexpected behavior. This vulnerability is also known in the literature
as 
"StateVariablesDefaultVisibility" \cite{tsankov_securify_2018}, 
 "Visibility level" \cite{tikhomirov_smartcheck_2018,zhang_soliditycheck_2019},
 "Unspecified visibility level" \cite{lu_neucheck_2019},
"Gain/Lose visibility" \cite{chapman_deviant_2019} or "SWC-108: State Variable Default Visibility" \cite{swc}.

\subsection*{\textbf{4.11 Redundant Functionality}}

Contracts that are written with redundant functionality increase code size and make maintainability difficult. In a simple scenario,  a programmer creates a function and later (by bad practices) ends up creating the same functionality again in a new function. He/she identifies a vulnerability in the new function and fixes it, but the old function with the defect is used by the caller. This vulnerability is also known in the literature as 
"Redundant refusal of payment"  \cite{zhang_soliditycheck_2019}, "Redundant fallback function" \cite{tikhomirov_smartcheck_2018}, or
"Unnecessary payable fallback function"  \cite{lu_neucheck_2019}.

\subsection*{\textbf{4.12 Shadowing}}

This category groups defects in which there are code elements (e.g., a function or a variable) with the same name, which can lead to erroneous and unexpected behavior.

\subsubsection*{4.12.1	Use of Same Variable or Function Name In Inherited Contract}

When using the same name as a local variable, which was previously declared by an inherited contract, the program loses the reference of the inherited variable, causing the local variable to assume the role of the other variable. This vulnerability is also known in the literature as 
"Shadowing state variables" \cite{tsankov_securify_2018}, 
"Shadow memory"  \cite{ashouri_etherolic_2020}, 
"Shadowing"  \cite{feist_slither_2019} or "SWC-119: Shadowing State Variables" \cite{swc}.

\subsubsection*{4.12.2	Variables or Functions Named After Reserved Words}

This bug occurs when creating variables named after keywords of the language itself. For example, in Solidity, creating a variable with the name \textit{now} conflicts with the function that returns the date and time. This vulnerability is also known in the literature as "ShadowedBuiltin" \cite{tsankov_securify_2018}.

\subsubsection*{4.12.3	Use of the Same Variable or Function Name In a Single Contract}

This vulnerability refers to cases where the same name is used for more than one variable or function inside the contract. This makes the program lose the reference of the variable of the class, assuming the variable of the function as its role. This vulnerability is also known in the literature as "ShadowedLocalVariable" \cite{tsankov_securify_2018}, "Redefined Variable" \cite{Hu2023} or
"Local variable shadowing" \cite{momeni_machine_2019}.

\subsection*{\textbf{4.13 Buffer Overflow}}

This category refers to overflow vulnerabilities (e.g., stack-based, heap-based) in which it is possible to write more data than what the buffer can hold, thus modifying memory areas outside the expected.

\subsubsection*{4.13.1 Stack-based Buffer Overflow}

The EVM keeps an execution stack that manages the execution of contracts. If an attacker is allowed to overflow this stack (e.g., by using specially crafted inputs), it can potentially overwrite control variables (e.g., timestamp or block number) and, for instance, gain unauthorized access to certain resources. This vulnerability is also known in the literature as "Stack size limited"  \cite{arganaraz_detection_2020}.

\subsubsection*{4.13.2 Write to Arbitrary Storage Location}

In solidity, arrays are stored as contiguous fixed-size slots. In the absence of a bounds verification, a malicious user could write data to a particular storage slot used to store the contract owner's address, which could be overwritten and then used to further harm the contract. This vulnerability is also known in the literature as "UnrestrictedWrite"  \cite{tsankov_securify_2018} ,
"Storage modification" \cite{krupp_teether_2018},"Arbitrary Write"	\cite{SMARTIAN_2021} or "SWC-124: Write to Arbitrary Storage Location" \cite{swc}.

\subsection*{\textbf{4.14 Use of Malicious Libraries}}

This defect refers to the use of third-party libraries containing malicious code. This vulnerability is also known in the literature as "Malicious libraries"  \cite{tikhomirov_smartcheck_2018},
"Unknown libraries" \cite{lu_neucheck_2019}, or
"Dynamic libraries"  \cite{andesta_testing_2020}.

\subsection*{\textbf{4.15 Typographical Error}}


This defect refers to single-digit errors made by programmers while typing source code, e.g., in logic or arithmetic operations. For example, for value assignment, a developer may type by mistake "$+=$" instead of "$=$" or may use "$-$" instead of "$+$" or "$--$" instead of "$++$" \cite{Hartel2019}. This vulnerability is also known in the literature as 
"Assignment operator replacement", "Binary operator replacement",
 "Unary operator replacement or deletion" \cite{Hartel2019}, or "SWC-129 Typographical Error"	\cite{swc}.

\subsection*{\hl{\textbf{5. Incorrect Control Flow}}}

This category groups a set of vulnerabilities that, if exploited, cause changes in the control flow of the program.




\subsection*{\textbf{5.1 Incorrect Sequencing of Behavior}}

This category gathers vulnerabilities that end up in a sequence of behaviors that are carried out in the wrong order, leading to unexpected results.

\subsubsection*{5.1.1 Incorrect Use of Event Blockchain variables for Time}

Contracts that rely on using block control information (i.e., timestamp, coinbase, number, difficulty, and gas limit) for sequential event control are vulnerable to tampering by the miner. This vulnerability is also known in the literature as 
"Block state dependence"  \cite{kalra_zeus_2018},
"Time restrictions"  \cite{arganaraz_detection_2020},
"Blockchain effects time dependency" \cite{zhang_ethploit_2020}, "Block State Dependency"	\cite{SMARTIAN_2021},
"Timestamp dependence" \cite{Ma2022, Hu2023,song_machine_2019,wang_contractward_2020,andesta_testing_2020,akca_solanalyser_2019,feng_precise_2019,luu_making_2016,nguyen_sfuzz_2020,ashraf_gasfuzzer_2020,Chen2020a,jiang_contractfuzzer_2018,tikhomirov_smartcheck_2018, zhang_soliditycheck_2019},
"Block number dependency"    \cite{Chen2020a,jiang_contractfuzzer_2018},
"BlockTimestamp" \cite{liao_soliaudit_2019},
"Time dependency" \cite{liao_soliaudit_2019},
"Race condition" \cite{ashouri_etherolic_2020}
"Block No. Dependency" \cite{ashraf_gasfuzzer_2020}, 
"Block number dependency" \cite{nguyen_sfuzz_2020}, 
"Event-ordering (EO) bugs" \cite{kolluri_exploiting_2019} or "SWC-116: Block values as a proxy for time" \cite{swc}.

\subsubsection*{5.1.2 Incorrect Function Call Order}

This defect refers to the creation of public functions that expect to be called in a certain sequence, originating unanticipated results whenever clients do not follow the right call order \cite{mavridou_designing_2018}. This vulnerability is also known in the literature , "Transaction-ordering dependence"   \cite{mavridou_designing_2018} or "SWC-114: Transaction Order Dependence" \cite{swc}.

\subsubsection*{5.1.3 Improper Locking}

This issue refers to the case where a contract assumes that all entities participating in a transaction must have the same credit balance before the contract operations can execute. If there are no adequate (e.g., wrong or even missing) locking mechanisms, an attacker can forcefully send credit to the other entity, which would cause the verification of the balance condition to never be met. Thus, the contract may become unusable or show unexpected behavior (or unexpected state changes). This vulnerability is also known in the literature as 
"IncorrectEquality" \cite{tsankov_securify_2018}, 
"Balance equality"  \cite{tikhomirov_smartcheck_2018, zhang_soliditycheck_2019},
"Strict equality"  \cite{lu_neucheck_2019},
"Strict Check for Balance" \cite{Chen2020a}, 
"Arbitrary sending of ether"  \cite{feist_slither_2019} or "SWC-132: Unexpected Ether balance" \cite{swc}.

\subsubsection*{5.1.4 Transfer Pre-Condition Dependent on Transaction Order}

In the case of this vulnerability, the order in which transactions are executed influence a pre-condition that guards the execution of the transfer. This influence may erroneously result in, for instance, a transaction not being executed at all. This defect is known in the literature as TODTransfer \cite{tsankov_securify_2018},
TOD \cite{SAILFISH_2022} or
Transaction Order Dependence  \cite{swc}.

\subsubsection*{5.1.5 Transfer Amount Dependent on Transaction Order }

This issue refers to the case where the value of the variable that stores or determines an amount of a digital asset (to be transferred) is modified before it is sent to the recipient due to transaction ordering within a block. The amount may be changed due to the effect of multiple transactions being grouped in a block and executed in a specific order having the effect of producing unexpected changes in the value being transferred. This vulnerability is also known in the literature as "TODAmount"  \cite{tsankov_securify_2018} ,
"TOD" \cite{liao_soliaudit_2019,wang_contractward_2020,SAILFISH_2022} or "SWC-114: Transaction Order Dependence" \cite{swc}.

\subsubsection*{5.1.6 Transfer Recipient Dependent on Transaction Order}

In the case of this defect, the transfer recipient is modified before the send event due to transaction ordering within a block. As an example, if the intended recipient address is stored as a storage variable and a transfer is to execute based on this address, there is a chance the address may be changed or overwritten by another transaction prior to the transfer. This vulnerability is also known in the literature as  "TODReceiver"  \cite{tsankov_securify_2018} , 
"Direct value transfer"  \cite{krupp_teether_2018}, "Transaction order dependence"  \cite{kalra_zeus_2018,Hu2023}, "Transaction-ordering dependence" \cite{song_machine_2019, bauer_semantic_2018, luu_making_2016,grieco_echidna_2020}, "TOD" \cite{SAILFISH_2022} or  "SWC-114: Transaction Order Dependence" \cite{swc}.

\subsubsection*{5.1.7 Exposed state variables}

This vulnerability refers to the case where a developer erroneously exposes a state variable, whose value may then be modified by an attacker so that this modification influences the execution of a certain contract operation. As an example, consider a contract that executes a credit transfer from one user to another and has a \texttt{require} statement for verifying that there is sufficient credit to conclude the operation. If the balance is stored as a public state variable, a malicious use could change its value so that the \texttt{require} is avoided allowing the user to run a transfer that exceeds the amount of credit actually held by the malicious user. This vulnerability is also known in the literature as "Vulnerable state" \cite{krupp_teether_2018}.

\subsection*{\textbf{5.2 Inadequate Input Validation}}

This group refers to defects involving the inadequate validation of functional conditions, which are requirements that a contract must meet so that it can operate correctly. Such conditions may offer protection against certain types of attacks or force certain business rules to be followed.

\subsubsection*{5.2.1 Improper Input Validation}

This type of problem occurs when an attacker calls a certain contract operation using invalid or malicious input data, capable of affecting the functioning of the contract due to the fact that either it does not validate the incoming inputs or validates them in an incorrect manner. For instance, in the context of Solidity, a Short Address Attack occurs when a contract receives less data than it was expecting, which leads the system to fill the missing bytes with zeros \cite{Chen2020a}. As a consequence, the behavior may become unexpected if the code assumes that the input data will comply with a certain length or format. This vulnerability is also known in the literature as 
"Invalid input data" \cite{Chen2020a,grieco_echidna_2020},
"Short address attack" \cite{ashouri_etherolic_2020},
"Short address" \cite{feng_precise_2019}, or
"Avoid non-existing address" \cite{chang_scompile_2019}.


\subsubsection*{5.2.2 Extraneous Input Validation}

In this particular case, the functional conditions of the contract are too strong and do not allow certain behaviors (which would be valid) to occur, making the contract unable to meet the requirements. This vulnerability is also known in the literature as "Requirement Violation"	\cite{SMARTIAN_2021} or 
"SWC-123 Requirement violation" \cite{swc}.






\subsection*{\hl{\textbf{6. Arithmetic Issues}}}

This category groups different vulnerabilities that share the outcome of resulting in arithmetic problems.

\subsection*{\textbf{6.1 Overflow and Underflow}}

This category refers to the use of operations (e.g., addition, subtraction) over values that result in a value that is less than (or greater than) the minimum values (or maximum value) that a variable can hold, which produces a value different from the correct result.

\subsubsection*{6.1.1 Integer Underflow}

This defect refers to operations over an Integer variable that result in a value that is less than the minimum value allowed by the Integer type. This vulnerability is also known in the literature as  "Integer Bug"	\cite{SMARTIAN_2021},
"Integer underflow"  \cite{smartpulse_2021,song_machine_2019,momeni_machine_2019,wang_contractward_2020,kalra_zeus_2018},
"Integer overflow/underflow"  \cite{wang_vultron_2019,lu_neucheck_2019,so_verismart_2020},
"Integer overflow and integer underflow" \cite{ayoade_smart_2019},
"Arithmetic bugs" \cite{torres_osiris_2018},
"Underflow" \cite{liao_soliaudit_2019},
"Integer overflow and underflow" \cite{nguyen_sfuzz_2020,ashouri_etherolic_2020},
"Overflow/Underflow" \cite{Ma2022,Cui2022,Hu2023,akca_solanalyser_2019}, 
"Arithmetic issues" \cite{andesta_testing_2020} or "SWC-101: Integer Overflow and Underflow" \cite{swc}.

\subsubsection*{6.1.2	Integer Overflow}

This defect refers to operations over an Integer variable that results in a value that is larger than the maximum value allowed by the Integer type. This vulnerability is also known in the literature as  "Integer Bug"	\cite{SMARTIAN_2021},
"Integer overflow" \cite{smartpulse_2021,wang_contractward_2020,kalra_zeus_2018},
 "Integer overflow/underflow" \cite{Ma2022,Cui2022,Hu2023,so_verismart_2020,wang_vultron_2019},
"Integer overflow and integer underflow" \cite{ayoade_smart_2019},
"Integer overflows" \cite{grieco_echidna_2020,grech_madmax_2020},
"Arithmetic Bugs" \cite{torres_osiris_2018}, 
"Overflow detector" \cite{gao_easyflow_2019}, 
"Overflow" \cite{liao_soliaudit_2019},
"Integer overflow and underflow"  \cite{nguyen_sfuzz_2020,ashouri_etherolic_2020},
"BatchOverflow" \cite{feng_precise_2019},
"Overflow/Underflow" \cite{akca_solanalyser_2019},
"Arithmetic issues"  \cite{andesta_testing_2020}, 
"Integer Overflow vulnerability"   \cite{song_machine_2019} or "SWC-101: Integer Overflow and Underflow" \cite{swc}.

\subsection*{\textbf{6.2 Division Bugs}}

This category groups issues related to erroneous division operations. 

\subsubsection*{6.2.1 Divide by Zero}

This issue refers to the attempt of a program to divide a value by zero. This vulnerability is also known in the literature as 
"Division-by-zero" \cite{so_verismart_2020},
"Arithmetic Bugs" \cite{torres_osiris_2018}, or
"Division by zero" \cite{akca_solanalyser_2019}.

\subsection*{6.2.2	Integer Division}

At the time of writing, a smart contract mainstream language like Solidity does not support floating point or decimal types. Thus, the remainder of a division operation is always lost. Developers may use fixed-point arithmetic and external libraries to handle this kind of operation. This vulnerability is also known in the literature as  "Numerical Precision Error" \cite{Cui2022},
"Integer division" \cite{tikhomirov_smartcheck_2018} ,
"Using fixed point number type" \cite{zhang_soliditycheck_2019} or "SWC-101: Integer Overflow and Underflow" \cite{swc}.

\subsection*{\textbf{6.3 Conversion Bugs}}

This category groups a set of vulnerabilities where there are issues related to the conversion between different datatypes.

\subsubsection*{6.3.1 Truncation Bugs}

This vulnerability refers to the case where a variable declared in a certain type is converted to a smaller type, which means that data is lost during the conversion process. This vulnerability is also known in the literature as "Truncation bugs"  \cite{torres_osiris_2018} or "SWC-101: Integer Overflow and Underflow" \cite{swc}.

\subsubsection*{6.3.2 Signedness Bugs}

The conversion of a signed integer type to an
unsigned type of the same width may change a negative value to a positive one (the opposite may also happen) \cite{torres_osiris_2018}. This vulnerability is also known in the literature as "Signedness bugs" \cite{torres_osiris_2018} or "SWC-101: Integer Overflow and Underflow" \cite{swc}.

\subsection*{\hl{\textbf{7. Improper Access Control}}}

This category groups a set of vulnerabilities that are strongly related to authentication or access control.

\subsection*{\textbf{7.1	Incorrect Authentication or Authorization}}

The smart contract fails to properly identify a client or determine its privileges, resulting in wrong access privileges for that particular client.

\subsubsection*{7.1.1 Wrong Caller Identification}

In Solidity, \texttt{tx.origin} allows obtaining the address of the account that initiated a transaction and \texttt{msg.sender} allows obtaining the address of the contract that has called the function being executed. The use of the \texttt{tx.origin} for access control may be a way of opening an entry point to a malicious user. A malicious user may create a contract that calls the vulnerable function (i.e., the one that uses \texttt{tx.origin} to check the identity of the caller). Thus, \texttt{msg.sender} will differ from \texttt{tx.origin}. In the case the vulnerable function uses \texttt{tx.origin} for access control, it will allow the user to perform actions it should not be able to. This vulnerability is also known in the literature as  "Transaction Origin Use"	\cite{SMARTIAN_2021},
"Transaction state dependence" \cite{kalra_zeus_2018},  
"Use of origin" \cite{Brent2018}, 
"TxOrigin" \cite{Hu2023,Li2023,akca_solanalyser_2019,liao_soliaudit_2019,tsankov_securify_2018},  "Tx.origin for authentication" \cite{zhang_soliditycheck_2019},  "Tx.origin" \cite{lu_neucheck_2019,tikhomirov_smartcheck_2018}, 
"Incorrect Check for Authorization" \cite{Chen2020a}, 
"Unprotected usage of tx.origin" \cite{momeni_machine_2019} or "SWC-115: Authorization through tx.origin" \cite{swc}.

\subsubsection*{7.1.2 Owner Manipulation}


This vulnerability allows an attacker to exploit some function or feature of the smart contract by manipulating the owner control variable. This allows the attacker to perform some kind of restricted operations \cite{zhang_ethploit_2020}. This vulnerability is also known in the literature as  "Missing Owner Check"	\cite{Cui2022}, "Unprotected Function" \cite{smartpulse_2021},
"Vulnerable access control" \cite{zhang_ethploit_2020},
"Access control" \cite{feng_precise_2019,lu_neucheck_2019}, or
"Tainted owner variable" \cite{brent_ethainter_2020}


\subsubsection*{7.1.3 Missing Verification for Program Termination}

This issue refers to the lack of a secure verification for terminating a published (deployed) contract, allowing an attacker to terminate it in an unauthorized manner. \texttt{Selfdestruct} is an EVM instruction that is able to nullify the bytecode of a deployed contract. When invoked, it stops the execution of the EVM, deletes the contract’s bytecode, and sends the remaining fund to a certain address. Access to this kind of function by non-authorized clients may result in security issues. This vulnerability is also known in the literature as  "Suicidal Contract"	\cite{SMARTIAN_2021},
"Unprotected Suicide"	\cite{Hu2023},
"Destroyable contract" \cite{Brent2018},
"SelfDestruct" \cite{liao_soliaudit_2019},
"Suicidal contracts" \cite{feist_slither_2019},
"Guard suicide" \cite{chang_scompile_2019},
"Unprotected usage of selfdestruct" \cite{momeni_machine_2019},
"Accessible selfdestruct" \cite{brent_ethainter_2020},
"Tainted selfdestruct" \cite{brent_ethainter_2020} or "SWC-106: Unprotected SELFDESTRUCT Instruction" \cite{swc}.

\subsection*{\textbf{7.2 Improper Protection of Sensitive Data}}

This category generally refers to the issues that result in the inability to protect sensitive information from non-authorized clients.

\subsubsection*{7.2.1 Exposed Private Data}

This issue refers to the cases in which contracts store unencrypted sensitive data in public blockchain transactions. Solidity, like other programming languages, support the \texttt{private} keyword that indicates that data is only accessible within the contract itself. However, in blockchain environments, marking a variable with \texttt{private} does not make it fully invisible to the outside world. Miners, which are responsible for validating transactions on the blockchain, can view the code of the contract and the value of its state variables \cite{zhang_soliditycheck_2019}. This vulnerability is also known in the literature as 
"Keeping Secrets" \cite{arganaraz_detection_2020},
"Exposed secret" \cite{zhang_ethploit_2020}, 
"Private modifier" \cite{lu_neucheck_2019,tikhomirov_smartcheck_2018,zhang_soliditycheck_2019} or "SWC-136: Unencrypted Private Data On-Chain"  \cite{swc}.

\subsubsection*{7.2.2 Dependency on External State Data (Unsolvable constraints of external critical state data)}

This vulnerability refers to the use of data that is not under control nor is generated by the contract (i.e., external critical state data). A malicious user may exploit this situation if such data determines the outcome of the execution of the contract. This vulnerability is also known in the literature as 
"Unsolvable constraints" \cite{zhang_ethploit_2020}.

\subsection*{\textbf{7.3 Cryptography Misuse}}

This category groups vulnerabilities that generally reflect misuse of cryptography mechanisms.

\subsubsection*{7.3.1 Incorrect Verification of Cryptographic Signature} 

This issue refers to the wrong verification of the authenticity and integrity of messages with the use of message signatures. As an example, a developer could develop a vulnerable contract that relies on a signature in a signed message hash for representing the earlier verification of previous messages. A client could generate a malicious message with a valid signature and include it in the hash. The contract then would validate the signature and update the hash, indicating that the message was processed. This vulnerability is also known in the literature as  "Missing Key Check"	\cite{Cui2022},
"SWC-117: Signature Malleability" \cite{swc}. 

\subsubsection*{7.3.2 Improper Check against Signature Replay Attacks}

This defect refers to a situation where a malicious client is able to obtain the message hash of a legitimate transaction and is allowed to use the same signature to impersonate the legitimate client and execute fraudulent transactions. This vulnerability is also known in the literature as 
"SWC-121: Missing Protection against Signature Replay Attacks" \cite{swc}.

\subsubsection*{7.3.3 Improper Authenticity Check}

In this case, a contract may tolerate off-chain signed messages instead of waiting for an on-chain signature. This is usually done with the goal of improving performance but may come at the expense of compromising the authenticity of the message. This vulnerability is also known in the literature as "Missing Signer Check"	\cite{Cui2022},
"SWC-122: Lack of proper signature verification" \cite{swc}.

\subsubsection*{7.3.4. Incorrect Argument Encoding}

This defect refers to the misuse of one-way hash functions (i.e., Solidity keccak256) namely in the incorrect encoding of the function arguments, which can result in a higher likelihood of hash collisions for different entries. This vulnerability is also known in the literature as
"Authorization" \cite{mavridou_designing_2018} ,
"Hash collision" \cite{lu_neucheck_2019} or "SWC-133: Hash Collisions With Multiple Variable Length Arguments" \cite{swc}.
\section{Discussion}
\label{sec:findings}

This section overviews the main characteristics of the taxonomy and maps our observations to state-of-the-art and industry practices. We conclude the section with a brief summary of the main aspects that contribute to the overall quality of the taxonomy.

\subsection{Mapping the taxonomy to the state of the art}

Table \ref{tab:tax-overview} summarizes the distribution of the number of vulnerabilities per each of the main categories present in our taxonomy. As we can see, the distribution is dominated by 'Bad Programming Practices \& Language Weaknesses', which account for almost half of the defects. Most of the remaining defects show relatively similar numbers among themselves.

\begin{table}[h!]
\centering
\caption{Vulnerability  distribution per taxonomy's main categories}
\label{tab:tax-overview}
\begin{tabular}{|l|c|}
\hline
\textbf{Category} & \textbf{\# Vulnerabilities} \\ \hline
Unsafe External Calls & 9 \\ \hline
Mishandled Events & 5 \\ \hline
Gas Depletion & 2 \\ \hline
Bad Programming Practices \& Language Weaknesses & 36 \\ \hline
Incorrect Control Flow & 9 \\ \hline
Arithmetic Issues & 6 \\ \hline
Improper Access Control & 9 \\ \hline
\end{tabular}
\end{table}

\begin{figure}[h!]
    \centering
    \includegraphics[scale=0.60]{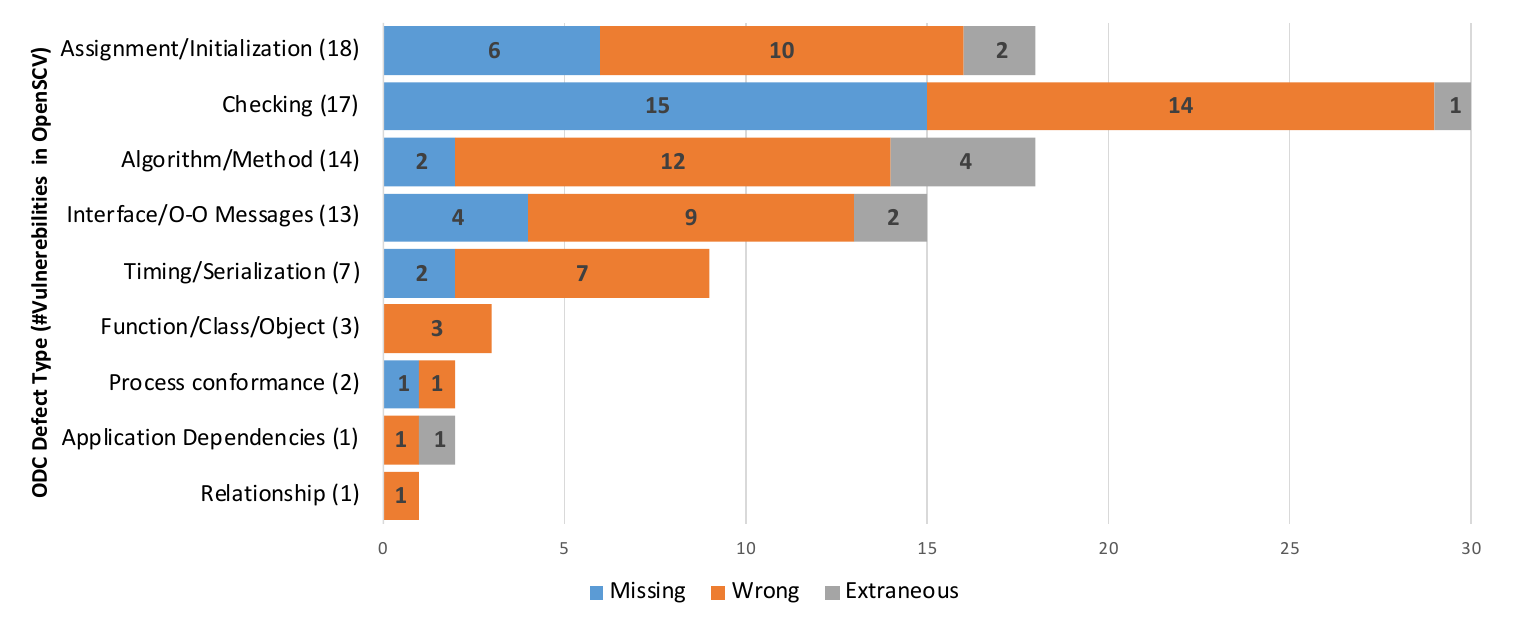}
    \caption{Vulnerabilities per ODC defect type and qualifier.}
    \label{fig:odc-qualifier}
\end{figure}

Figure \ref{fig:odc-qualifier} further characterizes the identified vulnerabilities, namely by identifying the different \textit{defect types} (in the y-axis) and specifying the number of OpenSCV vulnerabilities per each defect type (between parenthesis, in the y-axis). The plot then shows the prevalence of the \textit{qualifier} values. Notice that the sum of the qualifier values exceeds the vulnerability count between parenthesis in the y-axis, as a certain defect  may be associated with more than one qualifier.

As we can see in Figure \ref{fig:odc-qualifier}, the top \textit{defect types} fit in 'Assignment/Initialization' 'Checking', and 'Algorithm/Method', which is closely followed by 'Interface/O-O Messages'. The top three defect types account for nearly two-thirds of the defects, with the top four accounting for more than 80\% of the 76 vulnerabilities.

Table \ref{tab:combination} summarizes the \textit{qualifier} prevalence. It also shows the prevalence of qualifier combinations, which represent vulnerabilities whose correction may be related to more than one qualifier (e.g., a certain vulnerability may be due to a 'missing' or due to a 'wrong' 'assignment'). The 'wrong' qualifier is the most frequent one, followed by 'missing'. In terms of combinations, 'missing' and 'wrong' are the most frequent case.

\begin{table}[h!]
\centering
\caption{ODC qualifier distribution.}
\label{tab:combination}
\begin{tabular}{|l|l|l|l|}
\hline
\textbf{Qualifier} & \textbf{Extraneous} & \textbf{Missing} & \textbf{Wrong} \\ \hline
Extraneous         & 10                  & 1                & 3              \\ \hline
Missing            & 1                   & 30               & 19             \\ \hline
Wrong              & 3                   & 19               & 58             \\ \hline
\end{tabular}
\end{table}

\begin{figure}[h]
    \centering
   \includegraphics[scale=0.50]{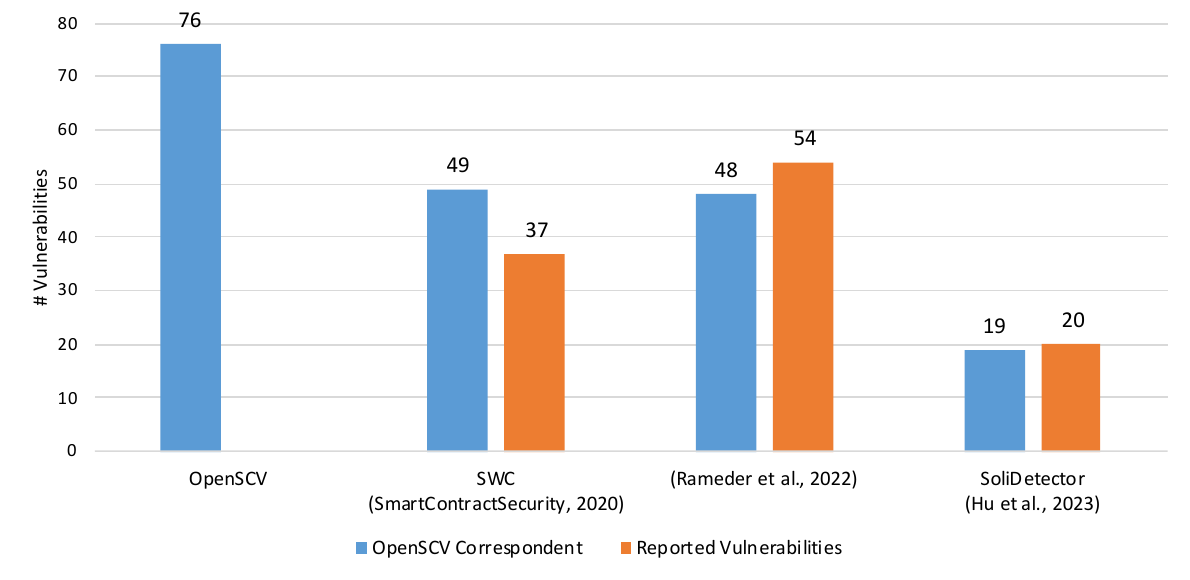}
    \caption{Mapping of the identified vulnerabilities to other classifications.}
    \label{fig:swc-dasp}
\end{figure}

We selected three different cases of classification schemes for comparison with OpenSCV. In particular, we selected SWC \cite{swc} for frequently appearing in the literature, we also selected the classification by \cite{Rameder2022} for being the most extensive one found in the state of the art, and we also selected the list of vulnerabilities used in \cite{Hu2023} for being the most recent vulnerability detection work in our list. Figure \ref{fig:swc-dasp} shows to what extent these classifications map the vulnerabilities identified in our taxonomy and overall shows their reach and practical limitations.

The first observation from Figure \ref{fig:swc-dasp} is that the number of distinct vulnerabilities currently captured by OpenSCV exceeds the remaining classifications in the plot. SWC has seen its 37 vulnerabilities unfold into 49 different cases. We also map all 48 vulnerabilities in \cite{Rameder2022}, in this case in a one-to-one mapping (notice that the work reports a total 54 defects, of which we excluded 6 that do not represent security vulnerabilities). Finally, 18 vulnerabilities in \cite{Hu2023} are mapped into 19 in OpenSCV (one of the vulnerabilities unfolds in two in OpenSCV). We must mention that \cite{Hu2023} actually identifies a total of 20 vulnerabilities, although, for two of them, there are no sufficient details to allow an understanding of what exactly the defect represents.

\begin{figure}[h]
    \centering\includegraphics[scale=0.70]{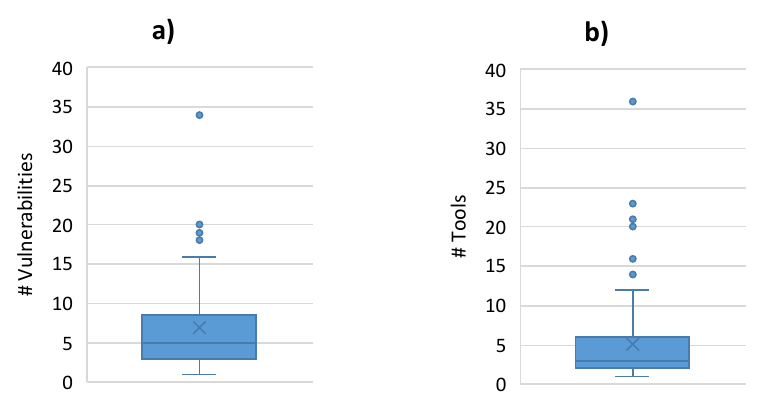}
    \caption{Announced detection capabilities of current tools: a) average number of vulnerabilities detected per tool; b) average number of tools per vulnerability.}
    \label{fig:tool-coverage}
\end{figure}

We now analyze the state of the practice (in terms of tools) by analyzing their announced vulnerability detection capabilities facing the identified vulnerabilities in our taxonomy. Figure \ref{fig:tool-coverage}.a) shows the practical distance of the 49 identified works in vulnerability detection to our current state of knowledge in what concerns smart contract vulnerabilities. Figure \ref{fig:tool-coverage}.b) shows, from the perspective of each individual vulnerability, how many tools are being designed to detect it. 

As we can see in Figure \ref{fig:tool-coverage}.a), current tools are being designed to detect in average 7  of the vulnerabilities in OpenSCV. If we consider the first quartile we see that tools are projected to detect from 3 to 8.5 different vulnerabilities. The best three vulnerability detection works (in terms of projected detection capabilities) are: Securify \cite{Tsankov2018} covering 44.8\% of OpenSCV vulnerabilities (34 out of 76); Neucheck \cite{lu_neucheck_2019} with 36\%, (20 out of 76) and; SoliDetector \cite{Hu2023} which is projected to detect 25\% of the vulnerabilities in OpenSCV (19 out of 76). We can see that, event in terms of design for detection, the room for improvement is huge and the plain combination of the different tools capabilities in itself is a clear possibility for the creation of a better tool. Figure \ref{fig:tool-coverage}.b) shows that a single vulnerability is being targeted, in average, by 5.1tools, with the first quartile  being between 2 and 6 tools.

\begin{figure}[h]
    \centering
    \includegraphics[scale=0.65]{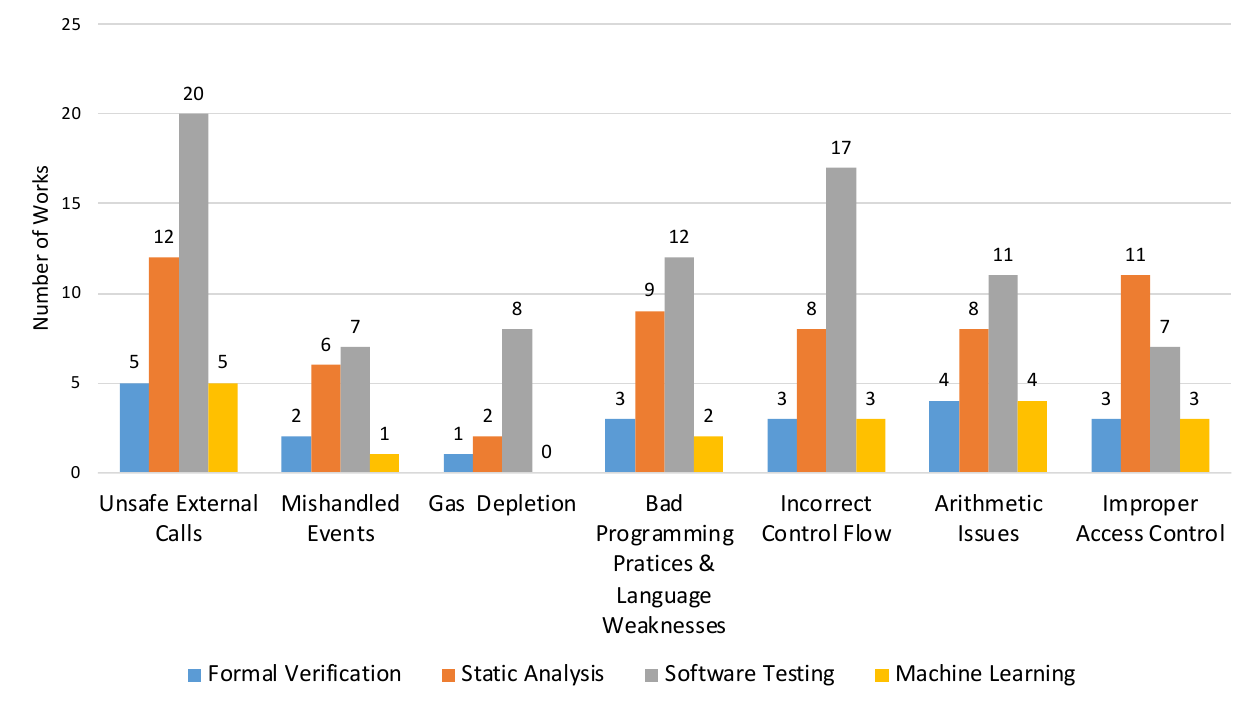}
    \caption{Announced detection capabilities for each of the taxonomy top categories.}
    \label{fig:toolVscategories}
\end{figure}

Figure \ref{fig:toolVscategories} shows the focus of the different classes of tools per each of the top categories in our taxonomy.

Figure \ref{fig:toolVscategories} shows that the category in which there is a larger focus from vulnerability detection approaches is "1. Unsafe External Calls", which is dominated by software testing approaches. Another aspect to mention is that, although "4. Bad Programming Practices \& Language Weaknesses" is the category that groups the largest set of vulnerabilities (i.e., 36 vulnerabilities), proportionally it is far from gathering the same attention as other categories (e.g., "6. Arithmetic Issues" is being targeted by 27 tools, although it groups only 6 types of vulnerabilities). The last aspect that is worthwhile mentioning is that software testing tends to be the most frequent technique across all categories of vulnerabilities, with the exception of "7. Improper Access Control", where static analysis has the lead. Indeed several vulnerabilities in this group easily fit the detection capabilities of static analysis techniques (e.g, cryptography misuse vulnerabilities). The top 5 vulnerabilities (in terms of presence in the different papers) are  "1.1.1 Unsafe Credit Changes" (36), "5.1.1 Incorrect Use of Event Blockchain Variables for Time" (28), "6.1.1 Integer Underflow" (21), "6.1.2 Integer Overflow" (21), "4.7.1 Unreachable Payable Function" (18). Detailed information and further data (e.g., code examples) can be quickly viewed at the OpenSCV website \cite{openscvSite}.

\subsection{Main contributors to the overall quality of the taxonomy}

We now summarize the main aspects which we believe are the main contributors to the general quality of the taxonomy, which we are making publicly available at \url{http://openscv.dei.uc.pt} \cite{openscvSite}. In terms of organization, we opted for a \textit{hierarchical structure}, as it may be useful from a defect prevention perspective. From a language designer's perspective, understanding that there is a certain group of defects that are related to, for instance, gas depletion may be helpful for designing effective protection mechanisms against those defects. Such mechanisms may share common strategies. 

A taxonomic structure of this kind allows setting \textit{homogeneous levels of abstraction} in an easier manner, which we iteratively tried to achieve, although this kind of goal is quite difficult as it should  be balanced with the number of items and overall tree complexity (and in some cases, due to the specificity of the problem, this may not even be possible). We tried to, as much as possible, \textit{reuse existing terminology} although many times we converged to the use of new terms (adapted from the literature), for clarity purposes. The required nomemclature adaptations integrated into our taxonomy were carried out mostly with the goal of making the items \textit{non-ambiguous} (and \textit{uniquely identifiable} also) and also fostering the \textit{determinism of the classification} process by clarifying the meaning of each vulnerability. We complemented this with the available information from DASP, SWC, \cite{Rameder2022}, and CWE, targeting to make the taxonomy further \textit{comprehensible} and \textit{non-ambiguous} (multiple perspectives will dissipate standing doubts, fostering \textit{repeatability}).

The taxonomy construction process involved the analysis of a relatively large number of papers, tools, and other classifications, with the main goal of fostering \textit{completeness} (i.e., good coverage), which in the end makes it also more \textit{useful} as we end up forming a unified view of the landscape of smart contract vulnerabilities. As previously mentioned, we found that the number of papers and respective vulnerabilities analyzed (i.e., an initial set of 357 vulnerabilities collected from 49 papers) was actually a main contributor to the overall quality of the taxonomy, with a few late additions becoming trivial to map. It is worthwhile mentioning that the created structure is non-ridig in the sense that we make it \textit{open to the community} and, in particular, open to community contributions, which can be carried out by submitting issue requests at the OpenSCV Github repository \cite{openscvGithub}.

\section{Threats to Validity}

This section discusses the main threats to the validity of this work. To minimize the chances of creating an \textit{incorrect structure or providing incorrect vulnerability information}, we formalized the taxonomy creation process, which was based on several quality criteria identified in the state of the art, and especially made use of several researchers (i.e., one Early Stage Researcher and 2 Experienced Researchers) who incrementally and iteratively built the taxonomy following a bottom-up approach. The process was enriched by establishing relations to other classifications in the blockchain context (i.e., SWC, DASP, \cite{Rameder2022}) and in a more general context (i.e., CWE). We also characterized each vulnerability using ODC and an example, which served also to minimize doubts and clear divergences among researchers. In addition, we provide the taxonomy as a live structure at \cite{openscvSite} supported by a Github repository \cite{openscvGithub} so that possible mistakes are corrected and also allow future updates, changes, and overall taxonomy evolution. 

We are aware that a classification or categorization scheme or \textit{a taxonomy may assume one of several possible forms}. We may have more or less main categories, we may have a deeper tree, the organization may or may not be hierarchical, and so on. While such diversity is acceptable (as long as the organization and individual items are correct), we opted to focus on the taxonomy creation process instead of on forcing a certain structure. For this purpose, we identified quality criteria, analyzed similar structures in the state of the art so that we could learn from possible mistakes and incorporate lessons learned by previous researchers. While the current structure is a proposal, we prepared it built to change and evolve, by opening it to the community and also by directly providing 'Request For Change' templates to facilitate changes or additions to the present form.

An important aspect is that the taxonomy creation process was guided by the research that was found during the analysis of the state of the art. Thus, we may have missed some relevant work in this context and, with time this gap may become greater. The fact that we were already aware of contributions coming from 3 areas: research on vulnerability classification, initiatives on vulnerability classification that are community-oriented, and research on vulnerability detection, allowed for a more efficient search, with which we believe captured representative research in this context. Despite this, and to mitigate possible gaps between the set of works considered to build OpenSCV and the set not captured during the collection of papers in this work, we prepared a supporting infrastructure to allow continuous update and evolution of OpenSCV. Thus, we will be able to capture and integrate new research in vulnerability detection that may bring in emerging smart contract vulnerabilities.
\section{Conclusion}

In this paper, we presented an open hierarchical taxonomy for smart contract vulnerabilities. The taxonomy is up-to-date according to the current state of the practice and is prepared to handle future modifications and evolution. To build the taxonomy we began by analyzing current vulnerability classification schemes for blockchain, we also analyzed announced detection capabilities of research on smart contract vulnerability detection, and we followed an iterative process to structure the taxonomy. We discussed the proposed taxonomy characteristics and coverage against the state of the practice. In particular, we analyzed the announced detection ability of current industry-level tools and mapped it to the different identified vulnerabilities. As future work, we plan on using this taxonomy as basis to define a benchmark for smart contract vulnerability detection tools.

\section*{Declarations}

\textbf{Conflict of Interests} The authors declare that they have no conflict of interests.

\vspace{0.5cm}

\noindent
\textbf{Data Availability} All data used to write this paper is available at Zenodo \cite{openscvZenodo} and linked to Github \cite{openscvGithub}. A visual representation of the taxonomy is available at \url{http://openscv.dei.uc.pt} \cite{openscvSite}.

\vspace{0.5cm}
\noindent
\textbf{Acknowledgments} This work is funded by the FCT - Foundation for Science and Technology, I.P./MCTES through national funds (PIDDAC), within the scope of CISUC R\&D Unit - UIDB/00326/2020 or project code UIDP/00326/2020; and by Project "Agenda Mobilizadora Sines Nexus". ref. No. 7113, supported by the Recovery and Resilience Plan (PRR) and by the European Funds Next Generation EU, following Notice No. 02/C05-i01/2022, Component 5 - Capitalization and Business Innovation - Mobilizing Agendas for Business Innovation.


\bibliographystyle{spbasic}
\bibliography{library}

\end{document}